\documentclass[review]{elsarticle}
\journal{Acta Materialia}
\usepackage{amsmath}

\usepackage{amsfonts}
\usepackage{amssymb}

\usepackage{comment}

\usepackage{setspace}
\usepackage{graphicx}
\usepackage{subcaption}
\usepackage{gensymb}

\usepackage[a4paper,width=150mm,top=25mm,bottom=25mm,bindingoffset=0mm]{geometry}
\usepackage{graphics}
\usepackage{multirow}
\usepackage{adjustbox}
\usepackage [english]{babel}
\usepackage [autostyle, english = american]{csquotes}
\MakeOuterQuote{"}
\usepackage[normalem]{ulem}
\useunder{\uline}{\ul}{}
\usepackage[table,xcdraw]{xcolor}
\biboptions{numbers,sort&compress}
\usepackage{xr-hyper}
\usepackage[colorlinks]{hyperref}
\hypersetup{
	colorlinks,
	linkcolor={red},
	filecolor={red}, 
	urlcolor={red},
	citecolor={blue}
}
\usepackage{xr}
\usepackage{xr-hyper}
\usepackage[colorlinks]{hyperref}
\begin{document}
	
\begin{frontmatter}
	\title{A high-throughput ab initio study of elemental segregation and cohesion at ferritic-iron grain boundaries}
	\author[MPSusMataffiliation]{Han Lin Mai}
	\author[USYDaffiliation]{Xiang-Yuan Cui}
	\author[BAMaffiliation]{Tilmann Hickel}
	\author[MPSusMataffiliation]{J\"{o}rg Neugebauer \corref{mycorrespondingauthor}}
	\author[USYDaffiliation]{Simon P. Ringer \corref{mycorrespondingauthor}}
	\cortext[mycorrespondingauthor]{Corresponding authors: neugebauer@mpie.de, simon.ringer@sydney.edu.au}
	\address[MPSusMataffiliation]{Computational Materials Design Department, Max Planck Institute for Sustainable Materials, Max-Planck-Straße 1, 40237  Germany}	
	\address[USYDaffiliation]{School of Aerospace, Mechanical and Mechatronic Engineering and Australian Centre for Microscopy and Microanalysis, The University of Sydney, 2006 New South Wales, Australia}
	\address[BAMaffiliation]{BAM Federal Institute for Materials Research and Testing, 12489 Berlin, Germany}
	\begin{abstract}
		Segregation of alloying elements and impurities at grain boundaries (GBs) critically influences material behavior by affecting cohesion. In this study, we present an \textit{ab initio} high-throughput evaluation of segregation energies and cohesive effects for all elements in the periodic table (Z = 1–92, H–U) across six model ferritic iron GBs using density functional theory (DFT). From these data, we construct comprehensive elemental maps for solute segregation tendencies and cohesion at GBs, providing guidance for segregation engineering. We systematically assess the cohesive effects of different elements in all segregating positions along multiple fracture paths with a quantum-chemistry bond-order method as well as a modified Rice-Wang theory of interfacial cohesion. The effects of segregants on the cohesion of GBs are shown to vary drastically as a function of site character, and hence their induced cohesive effects must be considered as a thermodynamic average over the spectral energy distribution. Thus, models that overlook these aspects may fail to accurately predict the impacts of varying alloying concentrations, thermal processing conditions, or GB types. The insights presented here, along with our accompanying dataset, are expected to advance our understanding of GB segregation in steels and other materials.
	\end{abstract}
\end{frontmatter}
\section{Introduction}
Segregation of solutes or impurities at grain boundaries (GB) in engineering alloys can dominate their macro-scale behaviour \cite{lejcekGrainBoundarySegregation2010}. This phenomena is particularly important in structural alloys, where a small presence of alloying elements or impurities can drastically alter their mechanical behaviours. Segregation phenomena can manifest in famously detrimental forms, playing a key role in the hydrogen embrittlement \cite{johnsonRemarkableChangesProduced1875, johnsonIIRemarkableChanges1875} and temper embrittlement of metallic alloys \cite{olefjordTemperEmbrittlement1978, zabilskiiTemperEmbrittlementStructural1987}, which can cause the sudden and catastrophic failure of critical components in industrial infrastructure \cite{kalderonSteamTurbineFailure1972}. However, the phenomena can also be leveraged to improve the behaviour of alloys \cite{watanabeApproachGrainBoundary1984, watanabeGrainBoundaryEngineering2011}. Small additions of an element to the composition can result in drastically improved mechanical behaviours of an alloy, a process known as micro-alloying, in which segregation phenomena play a major role. As such, understanding segregation is key to the principles-guided design of alloys.
\\\\
Due to the importance of GB segregation, significant time and efforts have been devoted to its study and modelling in steels \cite{gengEffectMoPd2000, gengInfluenceAlloyingAdditions2001,gengInfluenceAlloyingAdditions2001a, jinStudyInteractionSolutes2014, p.a.subramanyamInitioStudyCombined2019, huSoluteEffectsS32020, itoFirstprinciplesStudyGrain2019, itoElectronicOriginGrain2020, kholtobinaEffectAlloyingElements2021, jiangEffectsAlloyingElements2022, maiSegregationTransitionMetals2022,  maiPhosphorusTransitionMetal2023}. Steels remain one of the most important engineering alloys, but are susceptible to various segregation embrittlement mechanisms that result in intergranular failure. Although a significant body of simulation work already exists, there is still a lack of consistent high-throughput data that details the behaviours of elemental segregation and its effects on cohesion \cite{wagihGrainBoundarySegregation2024,scheiberHighThroughputFirstPrinciplesCalculations2024}. In particular, there exists scarce computed segregation and cohesion data for certain element groups, such as the lanthanides (rare earths), which have been of historical \cite{vahedThermodynamicsRareEarths1976, waudbyRareEarthAdditions1978} and recent \cite{wangEffectRareEarth2020, wangStudyApplicationRare2008} interest in steelmaking. Furthermore, many studies in the literature are limited to investigations of single model GBs [e.g. \cite{gengEffectMoPd2000, gengInfluenceAlloyingAdditions2001,gengInfluenceAlloyingAdditions2001a, jinStudyInteractionSolutes2014, p.a.subramanyamInitioStudyCombined2019, huSoluteEffectsS32020, itoFirstprinciplesStudyGrain2019, itoElectronicOriginGrain2020, jiangEffectsAlloyingElements2022, yangHighthroughputFirstprinciplesInvestigation2023}]. This is problematic, because it has been shown that even the same element may be calculated to have significantly different segregation behaviours and effects on cohesion, depending on the model GBs used \cite{maiUnderstandingHydrogenEmbrittlement2021, maiSegregationTransitionMetals2022, maiPhosphorusTransitionMetal2023}. Moreover, the details in prior studies in ferritic Fe GBs are rarely consistent, with some opting to study a single site at a chosen GB \cite{liEffectCrMo2014, huSoluteEffectsS32020, jiangEffectsAlloyingElements2022, yangHighthroughputFirstprinciplesInvestigation2023}, or with different relaxation/calculation schema \cite{gengInfluenceAlloyingAdditions2001a}. Naturally, these factors combined result in a narrow understanding of segregation phenomena, since such small and non-consistent datasets do not allow for convincing extrapolation of observations, and hence do not allow more general conclusions to be drawn.
\\\\
Importantly, in the study of interfacial segregation and cohesion, studies often calculate a cohesive quantity known as the work of separation in the framework of Rice-Wang theory \cite{riceEmbrittlementInterfacesSolute1989, riceDuctileBrittleBehaviour1974}. In the most common applications of this theory, effects on cohesion are calculated as a difference in the optimised (minimised/relaxed) surface segregation and the GB segregation energies of a solute. As a result, such studies tacitly assume that the separation plane occurs on the plane of the segregated solutes. As such, typical studies employing Rice-Wang calculations generally do not account for how a solute may affect fracture topography, i.e., they do not search for the weakest cleavage plane, which is fundamental in determining intergranular cohesion. Fernandez et al. have argued the importance of studying varying fracture surfaces in evaluating cohesion of GBs \cite{fernandezStatisticalPerspectiveEmbrittling2022}. A simple example where this generates inaccurate descriptions of how solutes affect fracture, is in the case of strongly cohesion enhancing solutes. In these cases, the weakest cleavage plane typically exists away from the planes on which cohesion enhancing solutes are located, requiring explicit searches in order to elucidate their true effect on the intergranular cohesion, i.e., the lower bound. However, such searches are rarely conducted in prior studies, thereby neglecting fracture topography completely, resulting in inaccurate, often overestimated effects on cohesion.
\\\\ 
The effects that solutes have on GB cohesion must fundamentally be dictated by the bonding, which must be different at differing site environments. As such, it has been proposed that solute-induced changes of interfacial cohesion may be understood instead by performing chemical bonding analysis \cite{mcmahonTheoryEmbrittlementSteels1978, briantChemistryGrainBoundary1990}. However, only few recent studies \cite{itoElectronicOriginGrain2020, maiSegregationTransitionMetals2022, maiPhosphorusTransitionMetal2023} have utilised such methods in comparison to those that have utilised Rice-Wang to study GB cohesion in steels \cite{jinStudyInteractionSolutes2014, p.a.subramanyamInitioStudyCombined2019, huSoluteEffectsS32020, itoFirstprinciplesStudyGrain2019, itoElectronicOriginGrain2020, jiangEffectsAlloyingElements2022, yangHighthroughputFirstprinciplesInvestigation2023, zhangGrainBoundaryAlloying2024}.
\\\\
GBs are known to contain a \textit{spectrum} of sites that bind segregants \cite{whiteSpectrumBindingEnergies1977, huberMachineLearningApproach2018, wagihGrainBoundarySegregation2020}. Sites that do not possess the maximum segregation binding, but nevertheless do bind segregants, therefore must also contribute to overall GB cohesion. To address this, Aksoy et al. proposed a combined formulation of site-occupancies (their segregation energies) with the spectrum of cohesive effects that are enacted by segregants \cite{aksoySpectrumEmbrittlingPotencies2021}. However, many studies of GB cohesion neglect to treat the interplay between the expected occupancies at various sites and the distinct effects that these segregants must enact at these different sites. There are almost no ab-initio calculations that account for this interplay of effects. Since the expected occupation of sites by a solute at GBs varies across temperatures, so must their effective induced effects on GB cohesion. As such, the \textit{spectral} nature of segregation and its effect on a segregant's effects on GB cohesion has largely been neglected in prior studies.
\\\\
Density functional theory (DFT) calculations are often utilised to study GB segregation phenomena. However, such calculations are generally costly and thus most studies have been limited to investigating ad-hoc segregation phenomena of a few specific solutes or model GBs. To circumvent the costs and other limitations associated with DFT studies, some studies use empirical interatomic potentials to study segregation phenomena over a wider portion of the grain boundary parameter space \cite{huberMachineLearningApproach2018, wagihGrainBoundarySegregation2020}. However, this too comes with its own problems, such as the accurate prediction of segregation phenomena compared to those computed with ab-initio accuracy \cite{wagihGrainBoundarySegregation2024}. This is common, since the differences in the details of the methodology of the empirical potential creation and the properties used to calibrate them often result in highly heterogeneous areas of competence in property-predictions. Few potentials have been systematically evaluated for accuracy in capturing segregation phenomena through benchmarking against DFT results, and fewer still are specifically calibrated for the study of solute segregation \cite{wagihGrainBoundarySegregation2024}. Other problems involve the lack of empirical potentials for large swathes of the chemical composition space, which also limits their use in a systematic investigations of segregation behaviour in a high-throughput manner.
\\\\
Due to the aforementioned challenges, a holistic and thorough understanding of solute segregation and cohesion still eludes us. To address this, the segregation of solutes must first be investigated in a high-throughput manner across a variety of GBs that span the GB parameter space. These calculations should be performed with ab-initio level accuracy, such that reliable descriptions of segregation behaviour are gained. The conclusions drawn on the elemental effects on GB cohesion should account for both the spectral nature of segregation as well as their varied effects in different sites. Only then can a more complete understanding of segregation behaviour be attained, and in turn, guide our ability to design alloys through segregation engineering at GBs. The purpose of our study is to provide a consistent high-throughput \textit{ab initio} evaluation of the segregation behaviours and cohesive effects at ferritic iron GBs for all naturally occurring elements in the periodic table in a single dataset, extract chemical trends, and account for the fundamentally spectral nature of segregation while doing so, by evaluating all available sites at the GBs. 
\section{Methodology}
\subsection{DFT calculation details}
We performed first principles calculations based on DFT using the projector augmented wave (PAW) method \cite{blochlProjectorAugmentedwaveMethod1994} as implemented in the Vienna \textit{Ab initio} Simulation Package (VASP) \cite{kresseEfficiencyAbinitioTotal1996, kresseEfficientIterativeSchemes1996}. Spin polarization was accounted for in all calculations performed in this study. We utilized the generalized gradient approximation (GGA) via the Perdew-Burke-Ernzerhof (PBE) functional \cite{perdewGeneralizedGradientApproximation1996}. The Brillouin-zone integrations for all GBs employed $\Gamma$-centred $\textbf{k}$-point meshes, with an energy cut-off of 400 eV for the plane wave basis set. A first order Methfessel-Paxton scheme with a smearing width of 0.2~eV was adopted for all calculations. The electronic minimisation convergence criterion was set to 1x$10^{-5}$~eV, and the relaxation calculations were deemed converged when atomic forces were below 0.01~eV/\AA. Justifications for our selections of the exchange-correlation functional, $\textbf{k}$-point meshes, plane wave energy cut-off and grain lengths were presented in our prior published work \cite{maiSegregationTransitionMetals2022}. Note that in some cases for segregation and surface calculations used in the cohesion analysis, the supplied \textbf{k}-point meshes do not allow for calculation convergence. In these cases, we fall back to a KSPACING=0.5 setting. Calculation post-processing was done with a mixture of the pymatgen \cite{ongPythonMaterialsGenomics2013} and pyiron \cite{janssenPyironIntegratedDevelopment2019} VASP scrapers. The pseudopotential files used, i.e., their specific VASP POTCAR filenames, are tabulated in Table \ref{SI:tab:bulk_props} of the S.I. 
\subsection{Grain boundary structures}
Six coincident site lattice (CSL) type bcc-Fe GBs were selected as our model GBs in this study. These were the $\Sigma3[110](1\bar{1}1)$, $\Sigma3[110](1\bar{1}2)$, $\Sigma5[001](210)$, $\Sigma5[001](310)$, $\Sigma9[110](2\bar{2}1)$ and $\Sigma11[110](3\bar{3}2)$ GBs. The  $\Sigma3[110](1\bar{1}1)$, $\Sigma5[001](210)$, $\Sigma5[001](310)$ and $\Sigma11[110](3\bar{3}2)$ GBs are generally used as representative high-angle CSL GBs, whereas the $\Sigma3[110](1\bar{1}2)$ is a low-energy twin boundary. The $\Sigma9[110](2\bar{2}1)$ is notable for containing an edge-dislocation core-like structure, which is typically found in small angle tilt-GBs \cite{bhattacharyaInitioPerspective1102014}. Details on the cells used to study the GBs, the \textbf{k}-point meshes utilised for each model, and their characteristic GB energies ($\gamma_\text{GB}$) and rigid works of separation (W$_\text{sep}^\text{RGS}$) are presented in Table \ref{tab:grainboundary_pureprops}. The cells for all but the non-orthorhombic $\Sigma9[110](2\bar{2}1)$ GB were constructed using the Atomic Simulation Environment (ASE) \cite{larsenAtomicSimulationEnvironment2017} python package. The $\Sigma9[110](2\bar{2}1)$ GB was generated using GrainBoundaryGenerator of the pymatgen python package \cite{ongPythonMaterialsGenomics2013}. The resultant relaxed POSCARs of these pure GBs are available in the S.I.
\begin{figure}[h!]
	\centering
	\includegraphics[height=8cm]{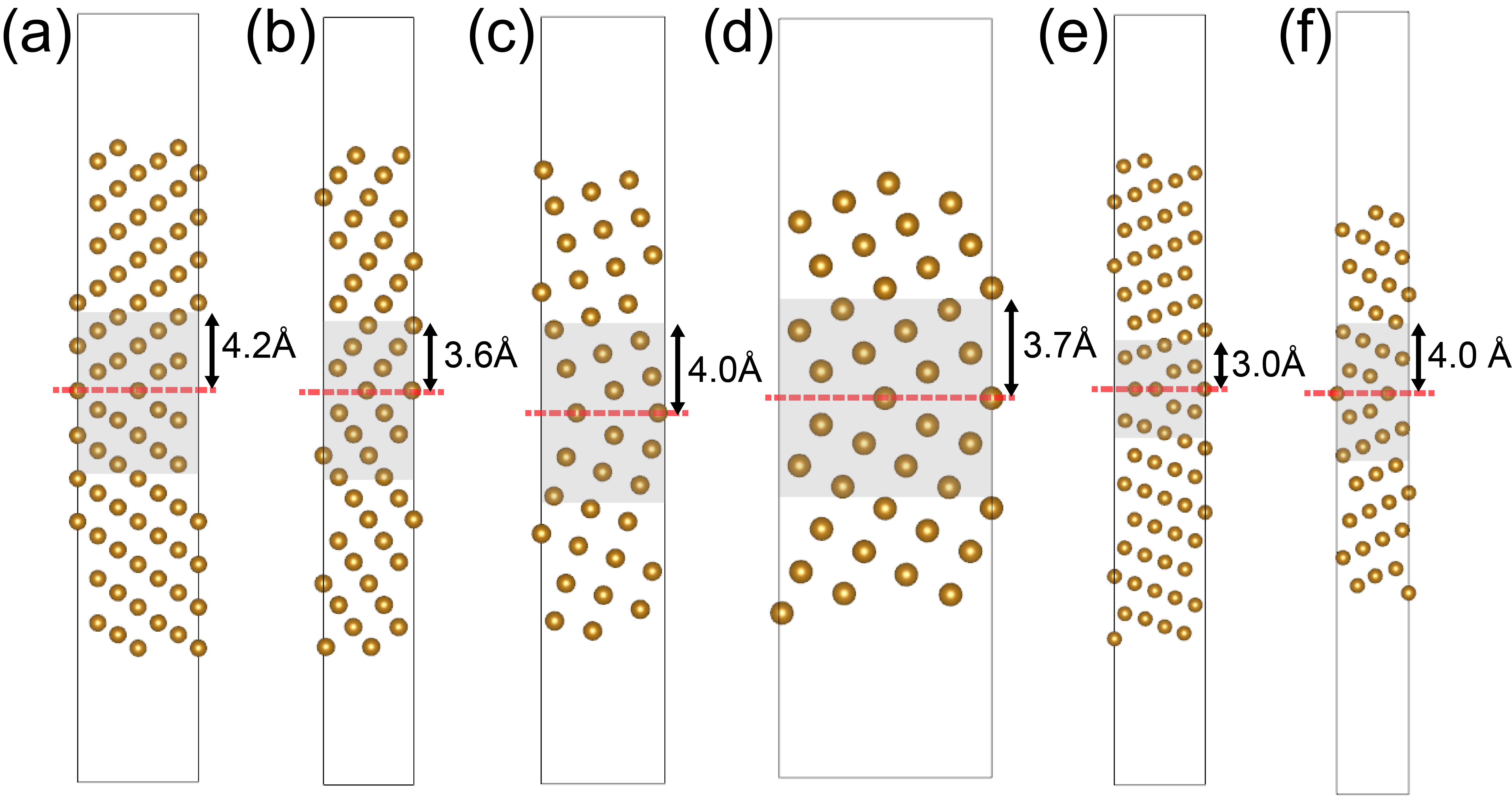}
	\caption{The atomic structures of the six coincident-site-lattice model GBs studied, visualised using VESTA\cite{mommaVESTAThreedimensionalVisualization2008}. The structures of the (a) $\Sigma3[110](1\bar{1}1)$,  (b) $\Sigma3[110](1\bar{1}2)$, (c) $\Sigma5[001](210)$, (d) $\Sigma5[001](310)$, (e) $\Sigma9[110](2\bar{2}1)$ and the (f) $\Sigma11[110](3\bar{3}2)$ GBs, respectively. The GB interface planes are highlighted by the dashed red lines. The shaded ranges indicate the range of studied sites for segregation. The dimensions of these cells are given in Table \ref{tab:grainboundary_pureprops}.}
	\label{fig:GBstructures}
\end{figure}
\begin{table}[h!]
	\centering
	\renewcommand{\arraystretch}{1.5}
	\makebox[\linewidth]{\begin{tabular}{ccccccccccc}
			\hline 
			System & 
			n$_\text{GB}$ & 
			\begin{tabular}[c]{@{}c@{}}a\\ (\AA)\end{tabular} & 
			\begin{tabular}[c]{@{}c@{}}b\\ (\AA)\end{tabular} & 
			\begin{tabular}[c]{@{}c@{}}c$_\text{GB}$\\ (\AA)\end{tabular} &
			\begin{tabular}[c]{@{}c@{}}Vacuum \\ (\AA)\end{tabular} & 
			\begin{tabular}[c]{@{}c@{}}Area \\ (\AA$^2$)\end{tabular} &
			\begin{tabular}[c]{@{}c@{}}$\gamma_\text{GB}$ \\ (J/m$^2$)\end{tabular} &
			\begin{tabular}[c]{@{}c@{}}W$_\text{sep}^\text{RGS}$ \\ (J/m$^2$)\end{tabular} &
			\begin{tabular}[c]{@{}c@{}}\textbf{k}-points \end{tabular} \\ \hline
			$\Sigma3[110](1\bar{1}1)$ & 72 & 4.005 & 6.937 & 28.740 & 15.40 & 27.78 & 1.58 & 4.19 & $6\times3\times1$\\ 
			$\Sigma3[110](1\bar{1}2)$ & 48 & 4.005 & 4.905 & 27.995 & 14.96 & 19.64 & 0.45 & 4.88 & $6\times6\times1$\\ 
			$\Sigma5[001](210)$ & 76 & 5.664 & 6.332 & 23.64 & 15.69 & 35.86 & 1.62 & 3.96 & $3\times3\times1$\\ 
			$\Sigma5[001](310)$ & 80 & 5.664 & 8.955 & 18.06 & 13.85 & 50.72 & 1.69 & 3.73 & $3\times3\times1$\\ 
			$\Sigma9[110](2\bar{2}1)$ & 68 &  4.005 & 6.332 & 23.640 & 19.32 & 24.06 & 1.75 & 4.22 & $6\times4\times1$  \\
			$\Sigma11[110](3\bar{3}2)$ & 42 & 4.005 & 4.696 & 24.770 & 15.13 & 18.81 & 1.45 & 4.27 & $6\times6\times1$ \\ \hline
	\end{tabular}}
	\caption{The sizes of the cells used in this study, the number of atoms, vacuum in the GB cells, the cross-sectional areas (for calculating the cohesive effects) and their corresponding GB energy ($\gamma_\text{GB}$), rigid work of separation are listed for all considered GBs.} 
	\label{tab:grainboundary_pureprops}
\end{table}
\subsection{Segregation energies}
\begin{figure}[h!]
	\centering
	\includegraphics[width=\linewidth]{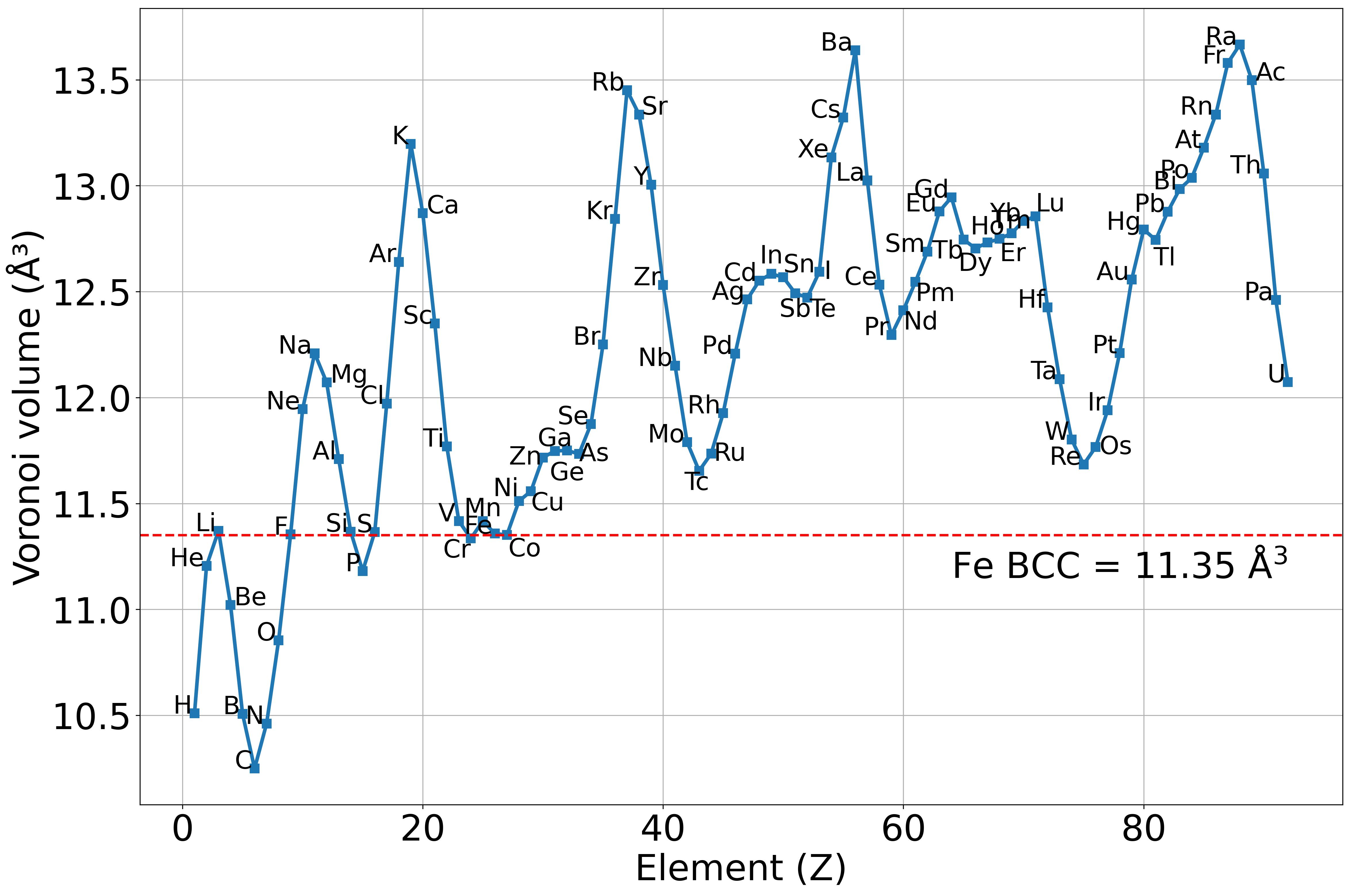}
	\caption{The Voronoi volumes occupied by solutes in a relaxed 128 atom BCC Fe cell, when a single Fe atom is substituted by the solute.}
	\label{fig:VorVolBulk}
\end{figure}

We computed segregation energy profiles at each GB for all elements Z=1$\to$92 (H$\to$U, exc. Yb) in the periodic table. The segregation energies were determined only via swaps of on-lattice Fe atoms in pure GBs with the solute atoms, which we deem as "substitutional" herein. The structures are then further relaxed until the force convergence criterion is met after the swaps. For the reference energy of the solute in the bulk, there are two cases, one in which the solute energetically prefers an on-lattice substitutional site, and the other case in which it prefers an interstitial position in the bulk. In the former case, the segregation energy of solute atom X (denoted $\text{E}_\text{seg}(\text{X})$) are calculated according to the formula:
\begin{equation}
	\text{E}_\text{seg}(\text{X}) = \text{E}_\text{GB}\big[(\text{n}-1)\text{Fe},\text{X}\big] - \text{E}_\text{GB} - \Big[ \text{E}_\text{Bulk}\big[(\text{m}-1)\text{Fe},\text{X}\big] - \text{E}_\text{Bulk} \Big]\quad.
\end{equation}
Here $ \text{E}_\text{GB}\big[(\text{n}-1)\text{Fe},\text{X}\big]$ is the energy of a GB supercell containing the substituted solute atom X, n is the number of atoms in the GB supercell,  $\text{E}_\text{Bulk}\big[(\text{m}-1)\text{Fe},\text{X}\big]$ is the energy of a bulk supercell containing substituted solute atom X, where m is the total number of atoms in the corresponding pure-Fe cell, and $\text{E}_\text{Bulk}$ the energy of the pure-Fe bulk supercell of the same size.
\\\\
In the cases of H, C, N, O, an interstitial position in the bulk is energetically preferred to the substitutional site. In this case, the formula is slightly changed due to the bulk reference energy calculation in the last term being different:
\begin{equation}
	\text{E}_\text{seg}(\text{X}) = \text{E}_\text{GB}\big[(\text{n}-1)\text{Fe},\text{X}\big] - \text{E}_\text{GB} - \Big[ \text{E}_\text{Bulk}\big[\text{m}\text{Fe},\text{X}\big] - \text{E}_\text{Bulk}/m + \text{E}_\text{Bulk}\Big]\quad.
\end{equation}
The possibility of solute atoms not lying directly on the on-lattice positions in the relaxed GB structures, often known as "interstitial" positions, were not considered in the segregation profiling performed in the present work. To justify this elimination of possible sites for most elements, we show that solutes which exhibit a size that would be indicative of possible preference towards interstitial positions at GB defects are rare across the periodic table. We hereto conduct a test that one may think of as analogous to the size-criterion of the Hume-Rothery rules, where the atomic size in the bulk is used to identify solutes which may possibly exhibit preference towards interstitial segregation at GBs. The difference in the occupied Voronoi volume of a substituted solute atom and a pure Fe atom in a relaxed 128-atom Fe-BCC bulk cell is used as the criterion for atomic size.
\\\\
We plot the Voronoi volume of the solute with respect to BCC-Fe across the elements in Fig. \ref{fig:VorVolBulk}, when substituted into a 128-atom BCC Fe cell. Here, only solutes which show significant (i.e., $>$0.4\AA$^3$) shrinkage in their occupied Voronoi volume are evaluated as possible interstitial GB segregants. According to this criterion, only six elements, H, Be, B, C, N, and O are interstitial segregants. Of similar size are He, Li, F, Si, P, S. We have previously demonstrated that P almost exclusively occupies substitutional sites in most of the GBs studied here \cite{maiPhosphorusTransitionMetal2023}. For S, it has been demonstrated that segregation in both interstitial and substitutional sites is possible \cite{yamaguchiDecohesionIronGrain2007, scheiberImpactSegregationEnergy2021}. As such, beyond the six elements of H, Be, B, C, N, O, which should prefer some interstitial sites at the GBs, most elements should prefer substitutional sites overall, at least in the highly symmetric CSL GBs considered here. We defer the study of the more complex case of these smaller solutes at interstitial segregation sites to a later work.
\subsection{Cohesion quantities}
Two GB-cohesion related quantities were considered in our evaluation of the effects of a solute on the interfacial binding strength. 
\begin{enumerate}
	\item The rigid grain separation (RGS) work of separation quantity, incorporating no atomic relaxations in the cleaved-surface calculation ($\rm{W}_{\rm{sep}}^{\rm{RGS}}$).
	\item The area-normalised summed bond-order (ANSBO) quantity.
\end{enumerate}

\subsubsection{Work of separation (RGS)}
The work of separation in the rigid-grain separation framework ($\text{W}_{\rm{sep}}^{\rm{RGS}}$) was calculated by:
\begin{equation}
	\text{W}_{\rm{sep}}^{\rm{RGS}} = (\text{E}_\text{GB-sep} - \text{E}_\text{{GB}}) / \text{A}\quad.
\end{equation}
Here, $\text{E}_\text{GB-sep}$ is the total energy of the cell containing a cleaved GB structure (with/without segregants) and $\text{E}_\text{{GB}}$ is the total energy of the corresponding non-cleaved structure. We emphasise that the atomic positions of the cleaved cell were \textit{not} relaxed. We did not compute works of separation with relaxed surfaces, as is common in studies utilising Rice-Wang theory. The atoms were cleaved with 6 \AA of vacuum separating the slabs. We have previously found that the cohesion quantities computed in the rigid methodology are more closely correlated with bond cleavage arguments \cite{maiSegregationTransitionMetals2022}, and hence should be more relevant to fracture-related phenomena overall, as has been argued similarly by others \cite{mcmahonTheoryEmbrittlementSteels1978, briantChemistryGrainBoundary1990}. 
\\\\
The cohesive effect that a solute has on a GB ($\eta_\text{RGS}$) may then be evaluated with:
\begin{equation}
	\eta_\text{RGS} = \text{W}_{\text{sep}}^{\text{seg}} - \text{W}_{\text{sep}}^{\text{pure}} \quad.
\end{equation}
Where $\text{W}_{\text{sep}}^{\text{pure}}$ and $\text{W}_{\text{sep}}^{\text{seg}}$ are the works of separation for the pure Fe GB and Fe GB containing a segregated solute, respectively. In this definition, positive values of $\eta_\text{RGS}$ indicate a solute has a cohesion enhancing effect, whereas negative values indicate that the solute has a detrimental on cohesion.
\subsubsection{Area-normalised summed bond orders}
The bonding-based area-normalised summed bond orders (ANSBO) quantity is defined as:
\begin{align}
	\centering
	\sum_{\substack{\text{frac}\\\text{path}}}\text{BO} = \sum_{i,j}^{i \neq j} \text{BO}[\text{X}_i,\text{X}&_j] + \frac{1}{2}\sum_{k,l}^{k \neq l} \text{BO}[\text{X}_k,\text{X}_l]\\
	\text{where}\ \text{X}_{1}(z) <&\ z_\text{CP} < \text{X}_{2}(z)
	\label{eqn:bondorder} \nonumber
\end{align}
\begin{equation}
	\text{ANSBO} = \sum_{\substack{\text{frac}\\\text{path}}}\text{BO} / \text{A}\quad.
\end{equation}
Here $\sum_{\substack{\text{frac}\\\text{path}}}\text{BO}$ is the chargemol \cite{manzIntroducingDDEC6Atomic2017} calculated summed DDEC6 bond orders of the electronic bonds participating in the cohesion of an arbitrary fracture path parallel to the GB plane, $\sum_{i,j}^{i \neq j}\text{BO}[\text{X}_i,\text{X}_j]$ is the bond order of the bond that exists between X$_i$, X$_j$, where X$_i$, X$_j$, X$_k$, X$_l$ are the atoms that are electronically bonded in sites $i$, $j$, $k$, $l$ respectively. The exact definition and derivation of the bond order is provided in the cited paper \cite{manzIntroducingDDEC6Atomic2017}. X$_1$, X$_2$ represent the $i,j$ and the $k,l$ pairings in any order. X$_i$ and X$_j$ atoms reside entirely within the supercell created (i.e., bonds exist wholly within the cell), whereas X$_k$ and X$_l$ represent atom pairs where only one of X$_k$ and X$_l$ resides in the cell (i.e., possess bonds passing outside of the original cell into a neighbouring image). $z_\text{CP}$ is the \textit{z} coordinate of the cleavage plane. Larger values of ANSBO indicate greater strength of interfacial cohesion. 
\\\\
Then, the cohesive effect that a solute has on a GB in the ANSBO framework ($\eta_\text{ANSBO}$) may be evaluated as:
\begin{equation}
	\eta_\text{ANSBO} = \text{ANSBO}_\text{seg} - \text{ANSBO}_\text{pure}\quad.
\end{equation}
Here $\text{ANSBO}_\text{seg}^{\text{pure}}$ and $\text{ANSBO}_\text{seg}$ are the ANSBOs for the pure Fe GB and Fe GB containing a segregated solute, respectively.
\\\\
Both quantities, the $\eta_\text{RGS}$ and $\eta_\text{ANSBO}$, were evaluated for all cleavage planes parallel to the GB plane, and within 3\ \AA\ of the segregated solutes. No cleavage quantities were evaluated between atoms on the same layer. Atoms were considered to be on the same layer where the distance in "z" direction is less than 0.1\ \AA. Beyond this criterion, no further assumptions were made about the weakest cleavage plane. They were explicitly searched for across all considered segregation sites in all GBs. Unless otherwise stated, all references to the cohesion effects of solutes, $\eta_\text{RGS}$ and $\eta_\text{ANSBO}$, are defined to be that of the smallest quantity calculated from the ensemble of separation planes considered.
\\\\
For purposes of direct comparison between the two frameworks, we write the cohesive effect of a segregant as a relative value to the cohesion measured at the original pure Fe interface, e.g. for ANSBO:
\begin{equation}
	R(\text{ANSBO}) = \text{ANSBO}_\text{seg}/\text{ANSBO}_\text{pure}\quad.
\end{equation}
and similarly for R($\text{W}_{\text{sep}}$).

\subsection{Thermodynamic effects}
We employed the White-Coghlan site-based extension \cite{whiteSpectrumBindingEnergies1977} of the Langmuir-McLean \cite{mcleanGrainBoundariesMetals1958} isotherm to calculate the expected occupancy at a site, $c_i$:
\begin{equation}
	c_i = \frac{c_B \times e^{\frac{-\rm{E}_{\text{seg}}^i}{k_{\text{B}} T}}}{1 - c_B + c_B \times e^{\frac{-\rm{E}_{\text{seg}}^i}{k_{\text{B}} T}}}\quad.
\end{equation}
Where $c_B$ is the alloying concentration of the solute in the bulk,  E$_{\text{seg}}^\text{i}$ is the segregation energy of the site $i$, k$_{\rm{B}}$ is the Boltzmann constant, and T is the equilibrium temperature. By summing the occupancies across the site-spectrum that exists at a GB, the interfacial coverage at the GB ($\bar{c}_{GB}$) is obtained:
\begin{equation}
	\bar{c}_{GB} = \sum_{i=1, \ldots, n} c_i = \sum_{i=1, \ldots, n} \frac{c_B \times e^{ \frac{-\rm{E}_{\text{seg}}^i}{k_B T} }}{1 - c_B + c_B e^{ \frac{-\rm{E}_{\text{seg}}^i}{k_B T}}} \quad.
	\label{eqn:white_coghlan_interfacialcoverage}
\end{equation}
Here, $i$ runs over the available sites at the GB, $1\to n$.
\\\\
Aksoy \textit{et al}. have proposed that the true effect that solutes have on cohesion must fundamentally be linked to their thermodynamic likelihood of occupying the segregation site \cite{aksoySpectrumEmbrittlingPotencies2021}. The effective cohesive effect provided by a specific site \textit{i} ( ($\epsilon_i$)) at a given temperature (\textit{T}) and bulk composition (c$_B$) (termed "embrittling estimator" in \cite{aksoySpectrumEmbrittlingPotencies2021}) is then computed by multiplying the cohesive effect of the solute at that site by the predicted occupancy:
\begin{equation}
	\epsilon_i = \eta_i(T, c_B) = min(\eta_i) * c_i\quad.
	\label{eqn:TempAdjCohesion_site}
\end{equation}
Where the cohesive effect at site \textit{i} ($\eta_i$)) is taken from the weakest plane evaluated for that site. In this study, $\eta$ takes two forms; the $\eta_\text{ANSBO}$ that is calculated in the DDEC6 bond-order based framework, and $\eta_\text{RGS}$, in the framework of rigid Rice-Wang theory. Then, to calculate the expected effect that an element has on GB cohesion ($\bar{\epsilon}_{GB}$) we sum these thermodynamically weighted effective cohesive effects over the sites present at the GB:
\begin{equation}
	\bar{\epsilon}_{GB} = \sum_{i=1, \ldots, n} \epsilon_i\quad.
	\label{eqn:TempAdjCohesion}
\end{equation}
This exploits the available data that we possess with respect to the full discretised values of segregation energies at GBs, and captures how GB cohesion is expected to vary with the differing amounts of solute occupation at the GB. Our approach is to calculate what is effectively a linear summation of the temperature and bulk composition adjusted lower-bound cohesive effect across all sites. The quantity of engineering interest is the lower-bound effect on cohesion, but these may not necessarily act on a single consistent plane between sites; but rather a collection. Since the effect that a solute has on GB cohesion must act on a collection of possible cleavage paths, the \textit{weakest} path may not necessarily occur on the same path for each site considered. In this manner, the effective cohesive effect ($\bar{\epsilon}_{GB}$), is a quantity that quantifies the cohesive effect of an ensemble of segregating sites on a collection of possible cleavage planes, specifically calculating the summed minimum lower bound of this action.  As such, we clarify that it is possible to calculate a value for the effective cohesive effect ($\bar{\epsilon}_{GB}$),that is larger than the pure-Fe GB cohesion quantity, which is a single-cleavage path quantity.
\\\\
Note that the temperature-adjusted cohesive effect is likely inaccurate in the non-dilute limit, where full occupation of multiple layers at the GB is expected. In this case, not only do solute-solute interactions come into play, but the linear scaling assumption of a solute on cohesive effect is also invalidated, as the bonds being replaced are no longer Fe-X bonds but rather X-X bonds, for which there is no data in our study. Nevertheless, it is instructive to observe the qualitative effect that solutes have on cohesion as a function of temperature, and these assumptions should not affect the conclusions that we draw from our analysis of the results.
\section{Results}
\subsection{Segregation energies}
\begin{figure}[h!]
	\centering
	\makebox[\textwidth][c]{%
		\begin{minipage}{\textwidth}
			\centering
			\includegraphics[width=\textwidth]{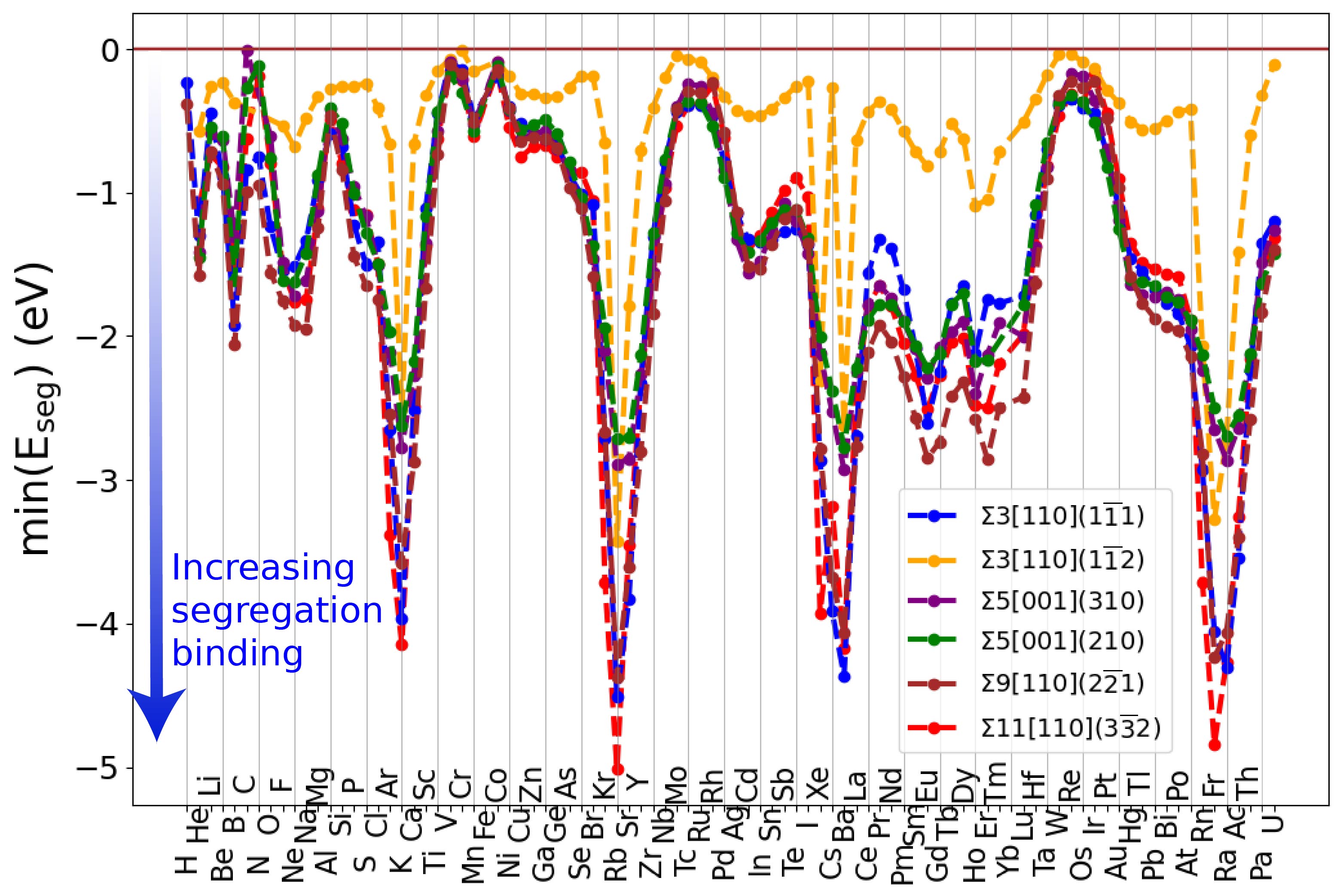}
			\caption{The minimum E${_\text{seg}}$ at each GB plotted against elemental number Z. These values are available in Table \ref{SI:tab:strongestbindingsites} in the S.I.}
			\label{fig:minEseg_vs_Z}
		\end{minipage}
	}
\end{figure}
\begin{figure}[h!]
	\makebox[0.7\textwidth][c]{%
	\begin{minipage}{\textwidth}
		\hspace{2.5cm}
		\includegraphics[width=\linewidth]{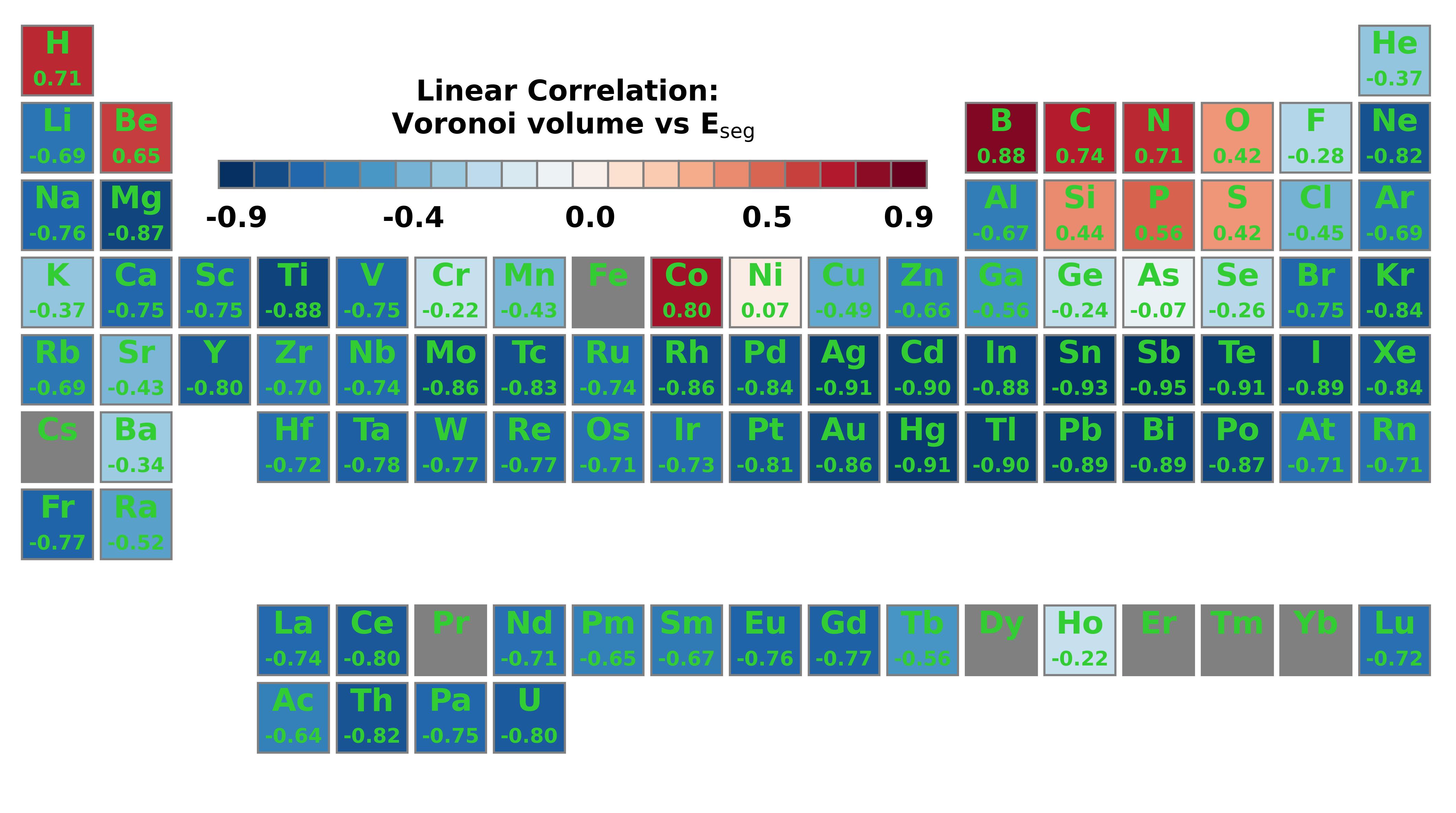}
	\end{minipage}
}
	\caption{The calculated linear (Pearson) correlation coefficients for the final relaxed segregation energies (E$_{\text{seg}}$) against the Voronoi volumes occupied by the solute at sites in a GB. The values in grey indicate elements for which this relationship was not calculated, as the segregation energy of Fe in Fe is ill-defined, and there is no data for Yb due to convergence issues.}
	\label{fig:Ptable_corrcoefs_EsegVorvol}
\end{figure}

The maximum segregation binding (i.e., minimum segregation energies) for each element are quite comparable across most of the GBs studied, with the exception of the twin boundary, the $\Sigma3[110](1\bar{1}2)$ GB. As such, the minimum segregation energy calculated at each GB is plotted in Figure \ref{fig:minEseg_vs_Z}, while the data for the segregation energies across all studied sites are presented in both tabular and heatmap forms in Table 	\ref{SI:tab:strongestbindingsites} and Figure \ref{SI:fig:heatmap_Eseg_vs_GBsite} respectively in the S.I. For a comparison of these segregation energies with other studies in the literature for the more commonly studied elements, we refer readers to Refs. \cite{maiSegregationTransitionMetals2022, maiPhosphorusTransitionMetal2023}. One remarkable feature in this data is the periodicity of deep segregation traps with respect to element number.
\\\\
\begin{figure}[h!]
	\centering
	\includegraphics[width=\linewidth]{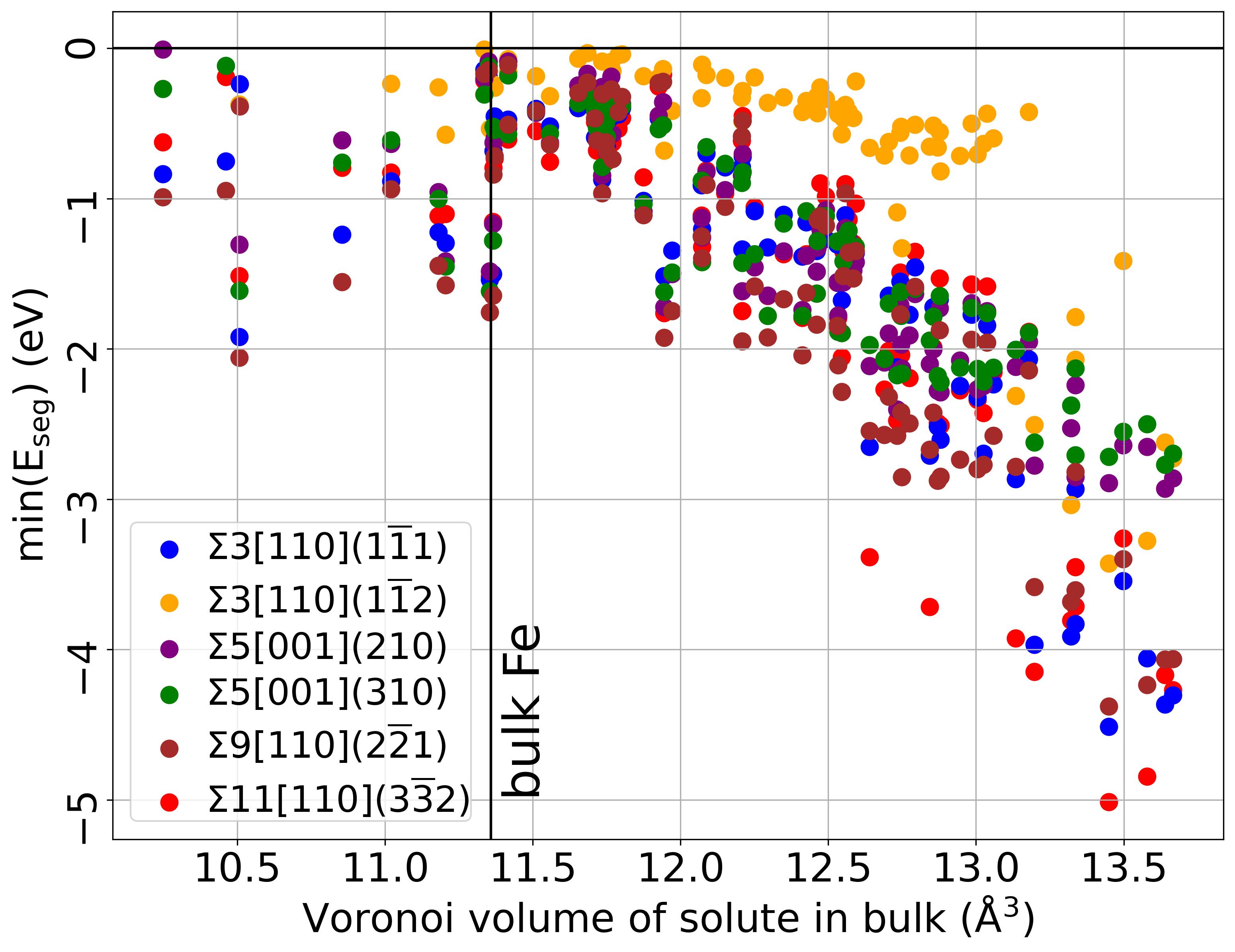}
	\caption{The strongest segregation tendencies at each GB, as indicated by minimum $\text{E}_\text{seg}$, is plotted against the Voronoi volume occupied by the solute in the bulk (in a 128-atom BCC cell). These values are available in Tables \ref{SI:tab:bulk_props}, \ref{SI:tab:strongestbindingsites} of the S.I.}
	\label{fig:Eseg_vs_VorVolBulk}
\end{figure}
To analyse this periodicity, we plot the minimum segregation energies at each GB for each solute also as a function of solute size in the bulk (see Methods, Fig. \ref{fig:VorVolBulk}) in Fig. \ref{fig:Eseg_vs_VorVolBulk}. We observe that the segregation energy overall decreases as a function of solute size. This indicates that elastic contributions/strain-related relaxation effects due to the solute size play a significant role in determining the maximum segregation binding that occurs in GBs. There is an exception in the case of the $\Sigma3[110](1\bar{1}2)$ twin-boundary, for which the relationship still holds, but is less pronounced. Some solutes, for example, Sr, K, Ra, Rb are very large and trigger giant relaxations at the twin $\Sigma3[110](1\bar{1}2)$ GB, in which case the solutes still satisfy the relationship found for more typical high energy GBs. However, these are the exception and not the norm. Note that there is some considerable scatter due to effects not directly attributable to strain; for example, for solutes very similar to Fe in size, such as Mn, Co, F, P, segregation binding is dominated by other chemical and/or magnetic effects, as one might expect. 
\\\\
To understand which elements have their segregation behaviour mostly governed by strain effects, we correlate their Voronoi volumes and calculated segregation energies in the final relaxed structures across their entire segregation profiles. The results of this analysis are visualized in periodic-table format in Fig. \ref{fig:Ptable_corrcoefs_EsegVorvol}. We have calculated this relationship for all elements using the linear (Pearson) correlation coefficient. We have also computed the rank (Spearman) correlation on the same data, without observing any noticeable differences from the results calculated by the linear correlation coefficients [see S.I. Fig. \ref{SI:fig:Ptable_spearmancoef_EsegVorvol}]. The large negative (positive) linear correlation coefficients indicate that the strength of segregation binding generally trends larger with increasing (decreasing) Voronoi volumes occupied by the solute. The dominance of darker shades of blue across most of the periodic table in both plots, confirms the overall trend of Fig. \ref{fig:Eseg_vs_VorVolBulk} that the segregation energy decreases as a function of solute size. 
\\\\

\begin{figure}[h!]
	\centering 
		\begin{subfigure}{\linewidth}
			\centering
			\includegraphics[width=0.8\linewidth]{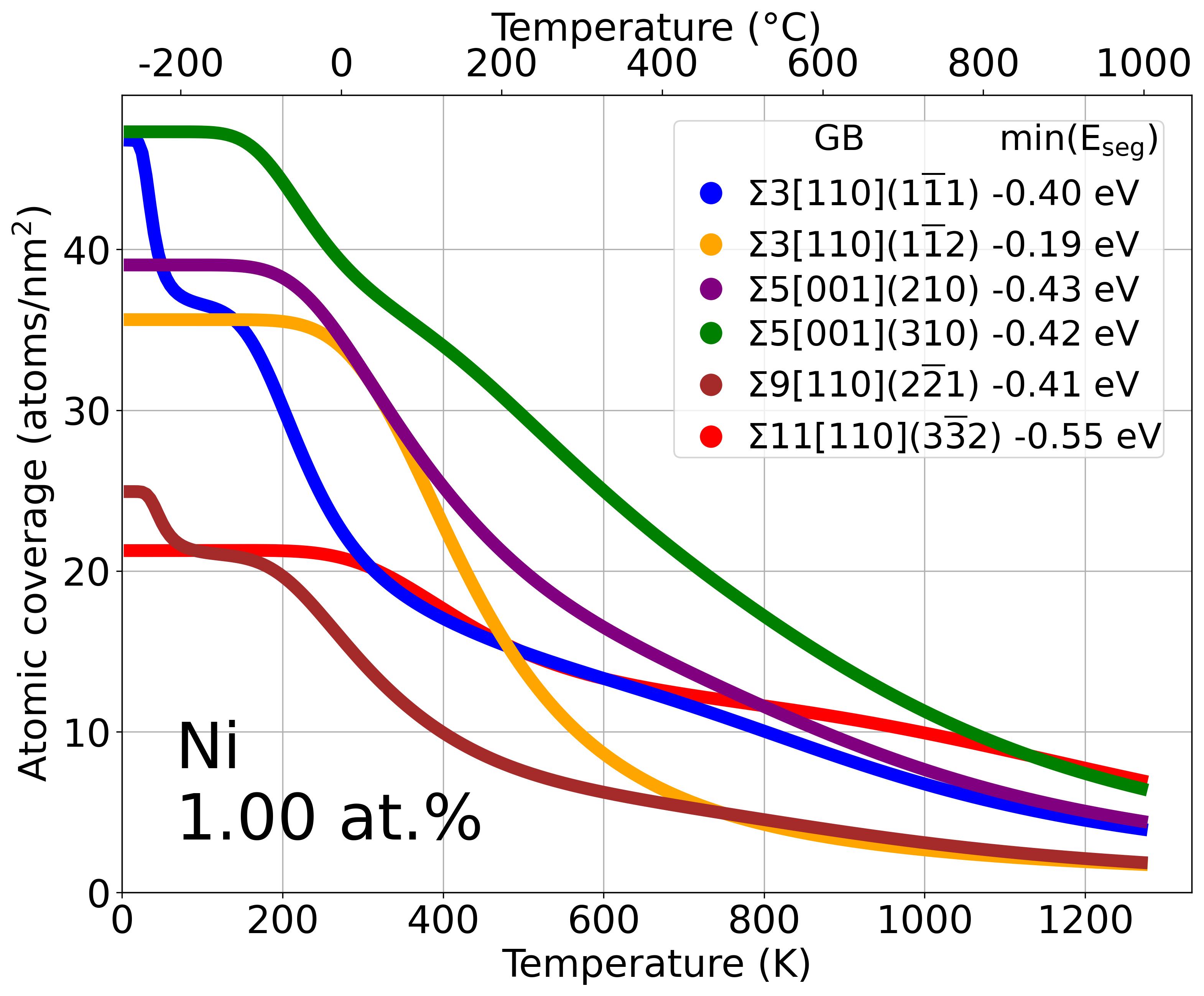}
			\caption{}
			\label{fig:WhiteCoghlan_Ni}
		\end{subfigure}
	\\ 
		\begin{subfigure}{\linewidth}
			\centering
			\includegraphics[width=0.8\linewidth]{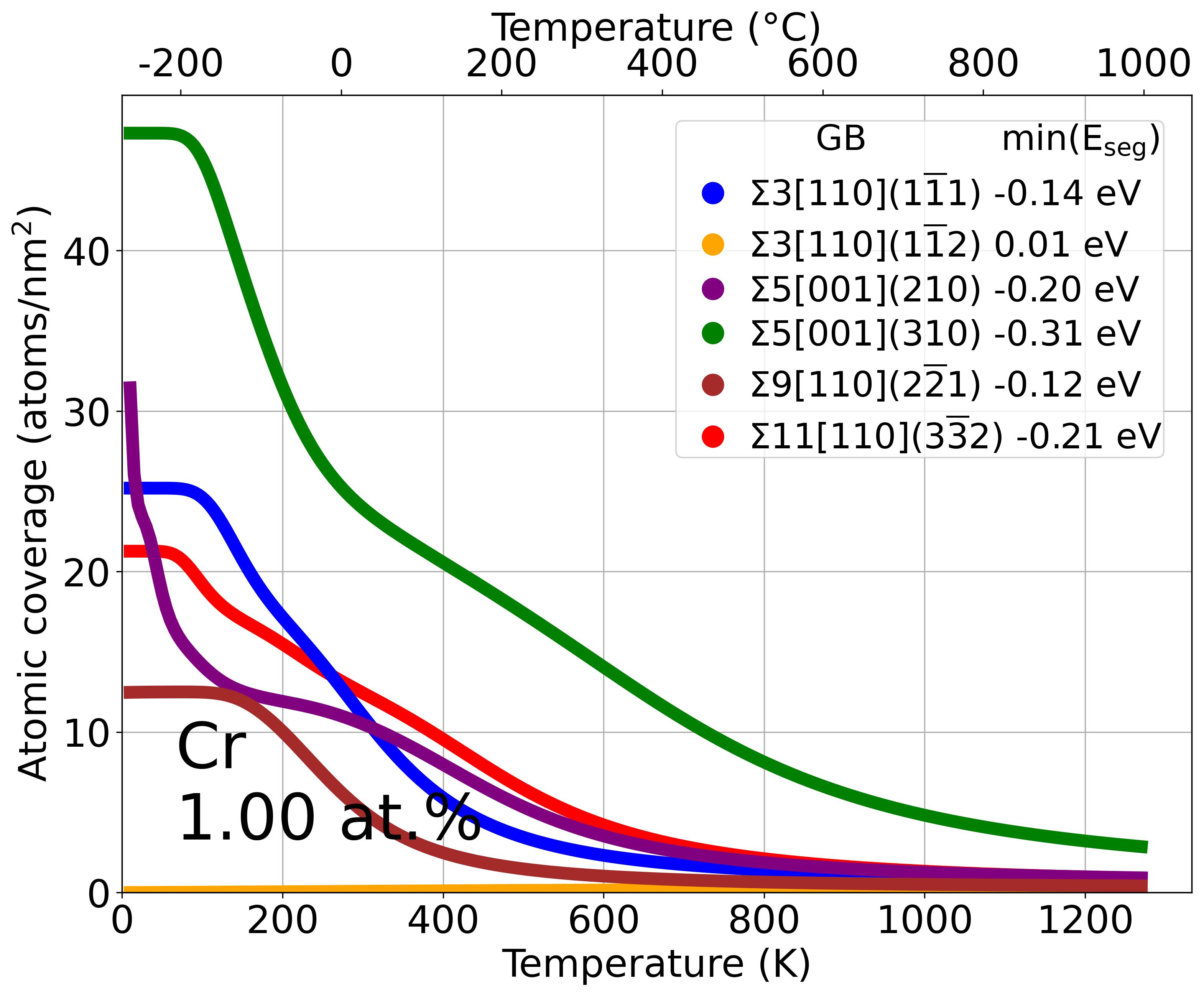}
			\caption{}
			\label{fig:WhiteCoghlan_Cr}
		\end{subfigure}
	\caption{The expected interfacial coverages in each model GB for a Fe-X binary alloy containing 1 at.\% of elements (\ref{fig:WhiteCoghlan_Ni}) Ni and (\ref{fig:WhiteCoghlan_Cr}) Cr. These were calculated using a White-Coghlan multi-site isotherm (see text). For all other elements, see the S.I.}
	\label{fig:WhiteCoghlan}
\end{figure}
McLean-style isotherm \cite{mcleanGrainBoundariesMetals1958} approaches assume that the segregation coverage (i.e. total amount of segregated solutes per-unit area interface) at GBs can be predicted by a single value, often prescribed to be the above-mentioned strongest segregation binding that exists at a GB. Instead, as White and Coghlan proposed \cite{whiteSpectrumBindingEnergies1977}, and others confirmed \cite{huberMachineLearningApproach2018, wagihGrainBoundarySegregation2020}, such approaches cannot accurately predict segregation coverages, since segregation \textit{spectra} exist at GBs instead. Although this fact is now well-known, actual visualisations of how the amounts predicted by simulated DFT \textit{spectra} compare against experiment are rare.
\\\\
In Figure \ref{fig:WhiteCoghlan} we plot the expected equilibrium interfacial coverage, as calculated with White-Coghlan, as a function of temperature of common steel alloying elements, Ni and Cr at Fe GBs at a bulk alloying composition of 1 at.\% across the model GBs studied. This illustrates that the expected amount of segregation at different types of GBs can change and even intersect at varying temperatures, due to the different spectrum of binding sites that exist at different GB structures. Therefore, prescribing single values of segregation energy to GBs, as is commonly done experimentally in Auger electron spectroscopy, result in area-averaged values that therefore cannot extrapolate interfacial coverages at GBs accurately. Such a method's predictions become increasingly unreliable when the thermal processing conditions or the precise mix of alloying elements depart from those in the initial experiments used to generate the fitted parameters. 
\\\\
In the following we demonstrate why approaches that do not treat the spectral nature of segregation, e.g., single value parameterisations of the phenomena, can fail to predict actual segregation coverage. We illustrate two cases in Fig. \ref{fig:TemperatureAdjSegregation}. First, the White-Coghlan predicted coverages, which account for the spectral nature of segregation at GBs, are plotted against the computed segregation energies at the strongest binding sites at GBs, in Fig. \ref{fig:minEseg_vs_WhiteCoghlanAmount}. The predicted coverage is generated at conditions of 0.1 at.\% alloying concentration at 300 K. Second, we simulate how area-averaged segregation energies, as one would measure from experimental techniques, i.e.,  by Auger electron spectroscopy, vary as a function of temperature due to the spectral nature of segregation, in Fig. \ref{fig:TempEffectiveSegregationEnergy}. The example taken in this case is for Nb, Mo segregation at the 0.087 at.\%/0.176 at.\% bulk alloying concentrations studied in the experiments in Ref. \cite{maruyamaInteractionSoluteNiobium2003}. This is achieved by taking the effective mean occupation ($c_{\text{GB}}$) calculated by White-Coghlan at the GB (Equation \ref{eqn:white_coghlan_interfacialcoverage}) and rearranging the equation to then solve for the effective segregation energy ($\bar{E}_{\text{seg}}$):
\begin{equation}
\bar{E}_{\text{seg}} = -k_B T \ln\left(\frac{c_{\text{GB}} (1 - c_B)}{(c_{\text{GB}} - 1) c_B}\right)\quad.
\end{equation}
In Figure \ref{fig:SiteEsegDOS_LangmuirDistribution_Nb} we plot the distribution of sites that are available at each GB considered against their probability to be occupied according to McLean isotherm theory. As can be seen, the distributions vary wildly between GBs, leading to different amounts of segregation at varying temperatures at different GBs.
\\\\
Note that in Fig. \ref{fig:minEseg_vs_WhiteCoghlanAmount} we do not plot the coverages for elements which have E$_\text{seg}<-1$~eV maximum segregation binding due to the their susceptibility to induce massive structural relaxations [see S.I. of \cite{maiPhosphorusTransitionMetal2023}] that result in a shifting of the GB plane/morphing of the local structure into equivalent final structures, despite starting at structurally different sites. In these cases, the segregation profiles generated are not physically meaningful, since the shifting renders substantial changes in the sites available to other solutes at the GB. Therefore, these energies cannot be used in our White-Coghlan analysis to generate realistic predictions on the amount of segregation that may occur at a GB. Nevertheless, as a series of companion plots to Fig. \ref{fig:minEseg_vs_Z}, the predicted coverages at varying alloying concentrations and temperatures are plotted against element number in Figs. \ref{SI:fig:WhiteCoghlan_vs_Z_300K_0.1at}-\ref{SI:fig:WhiteCoghlan_vs_Z_900K_1at} in the S.I. for interested readers.

\begin{figure}[h!]
	\centering
	\begin{subfigure}{1.0\linewidth}
		\centering
		\includegraphics[width=\linewidth]{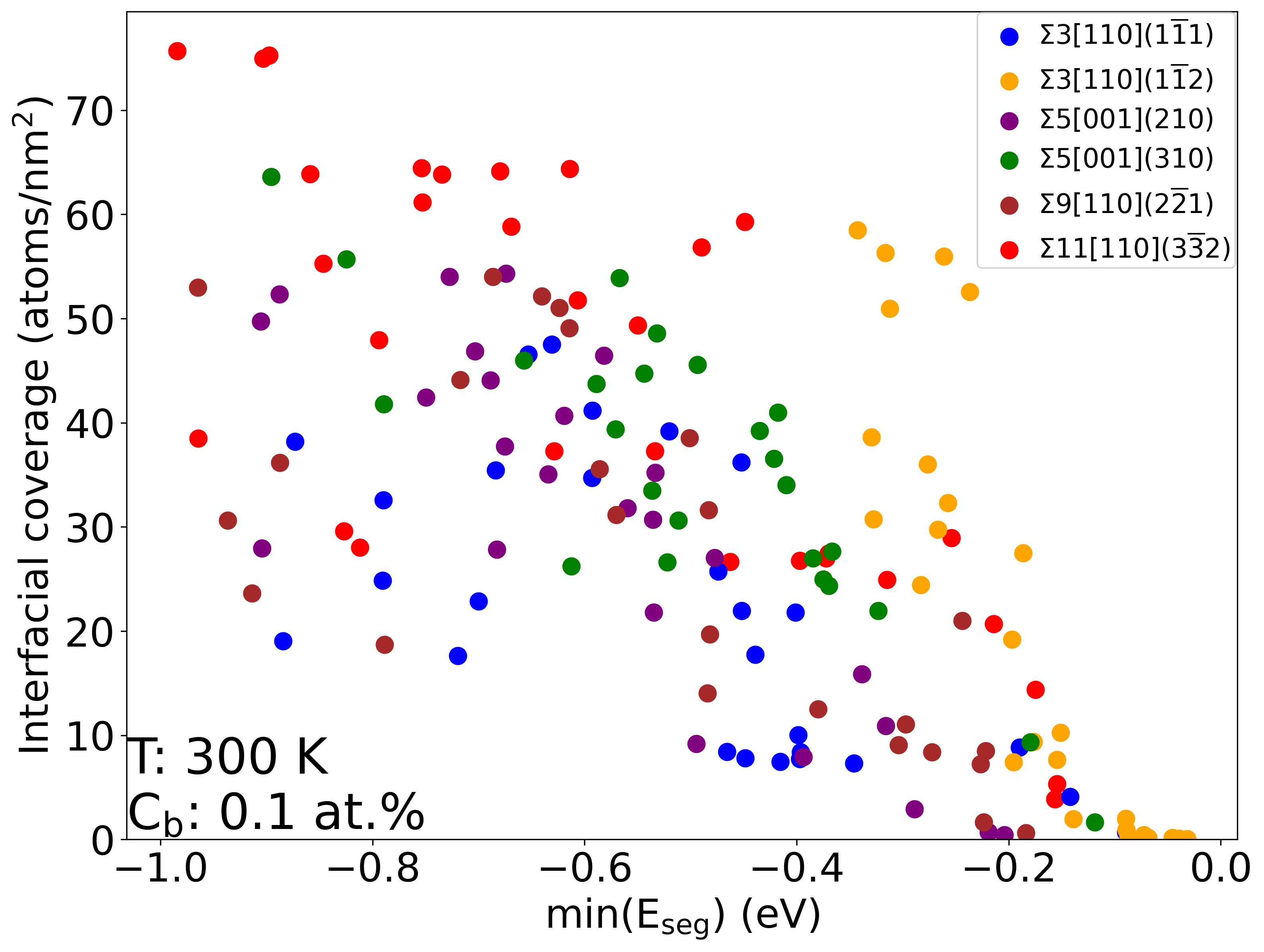}
		\caption{}
		\label{fig:minEseg_vs_WhiteCoghlanAmount}
	\end{subfigure}
	\\
	\begin{subfigure}{1.0\linewidth}
		\centering
		\includegraphics[width=\linewidth]{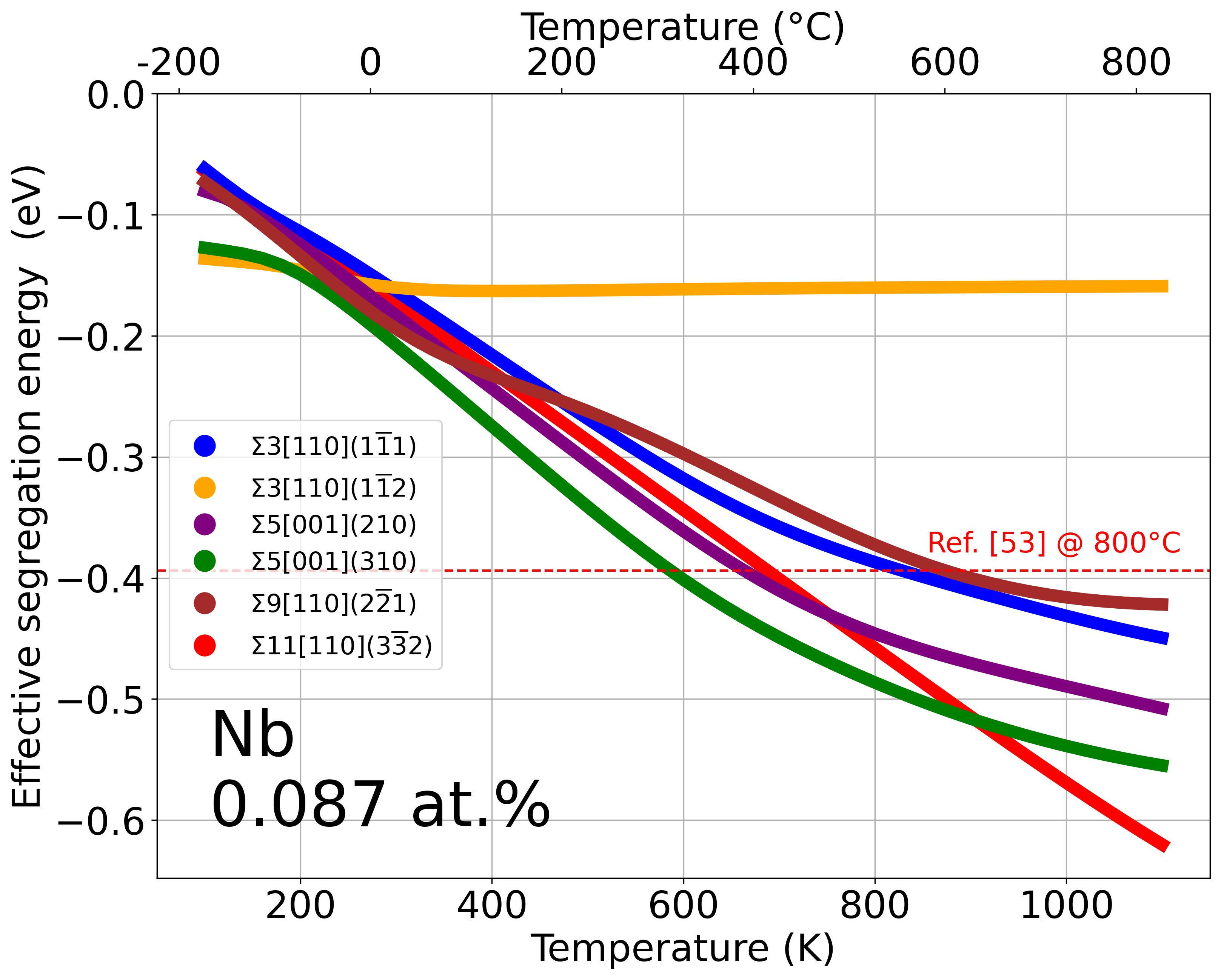}
		\caption{}
		\label{fig:TempEffectiveSegregationEnergy}
	\end{subfigure}
\end{figure}
\begin{figure}[h!]
	\ContinuedFloat
	\centering
	\begin{subfigure}{1.0\linewidth}
		\centering
		\includegraphics[width=\linewidth]{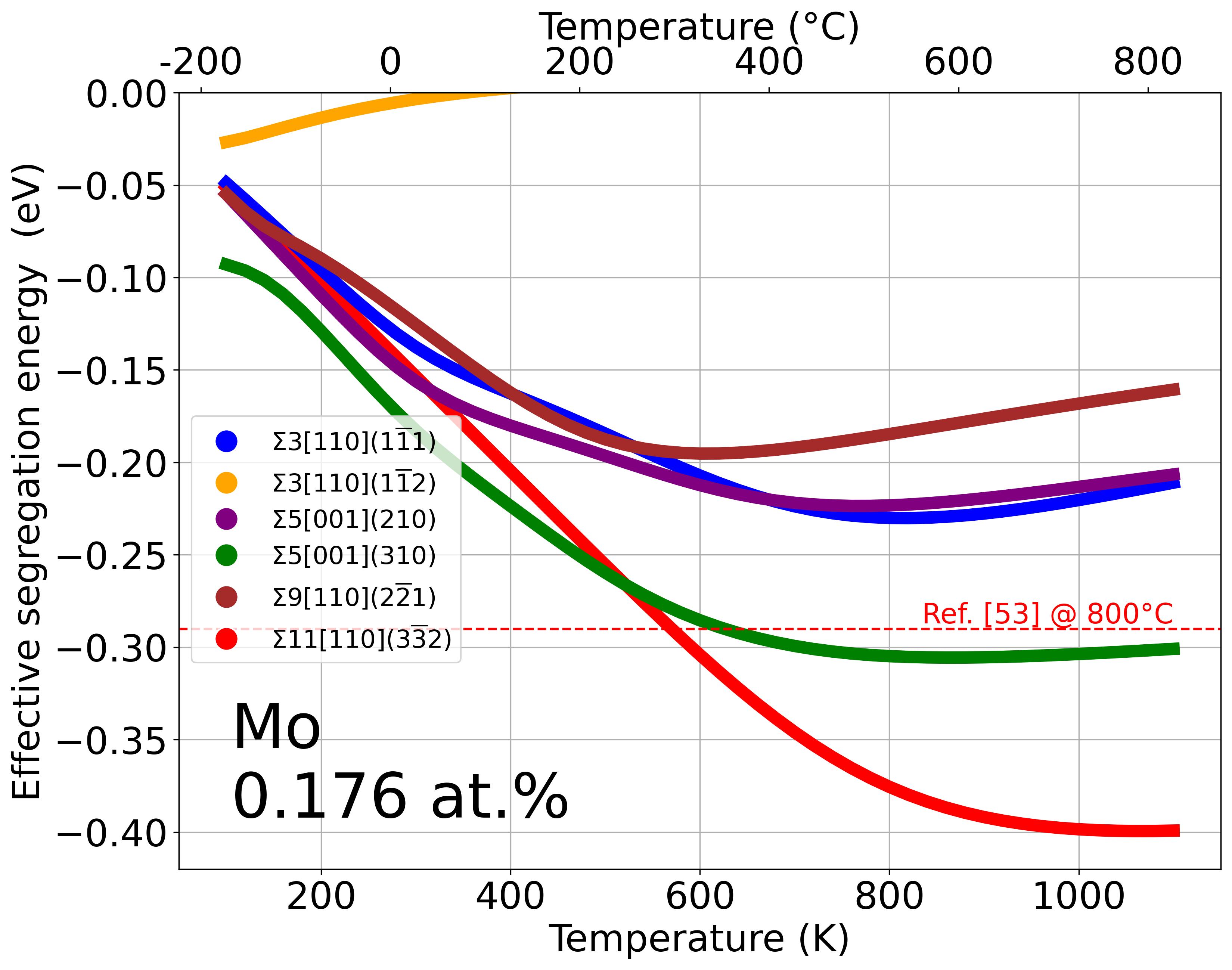}
		\caption{}
		\label{fig:TempEffectiveSegregationEnergy_Mo}
	\end{subfigure}
	\begin{subfigure}{1.0\linewidth}
		\centering
		\includegraphics[width=\linewidth]{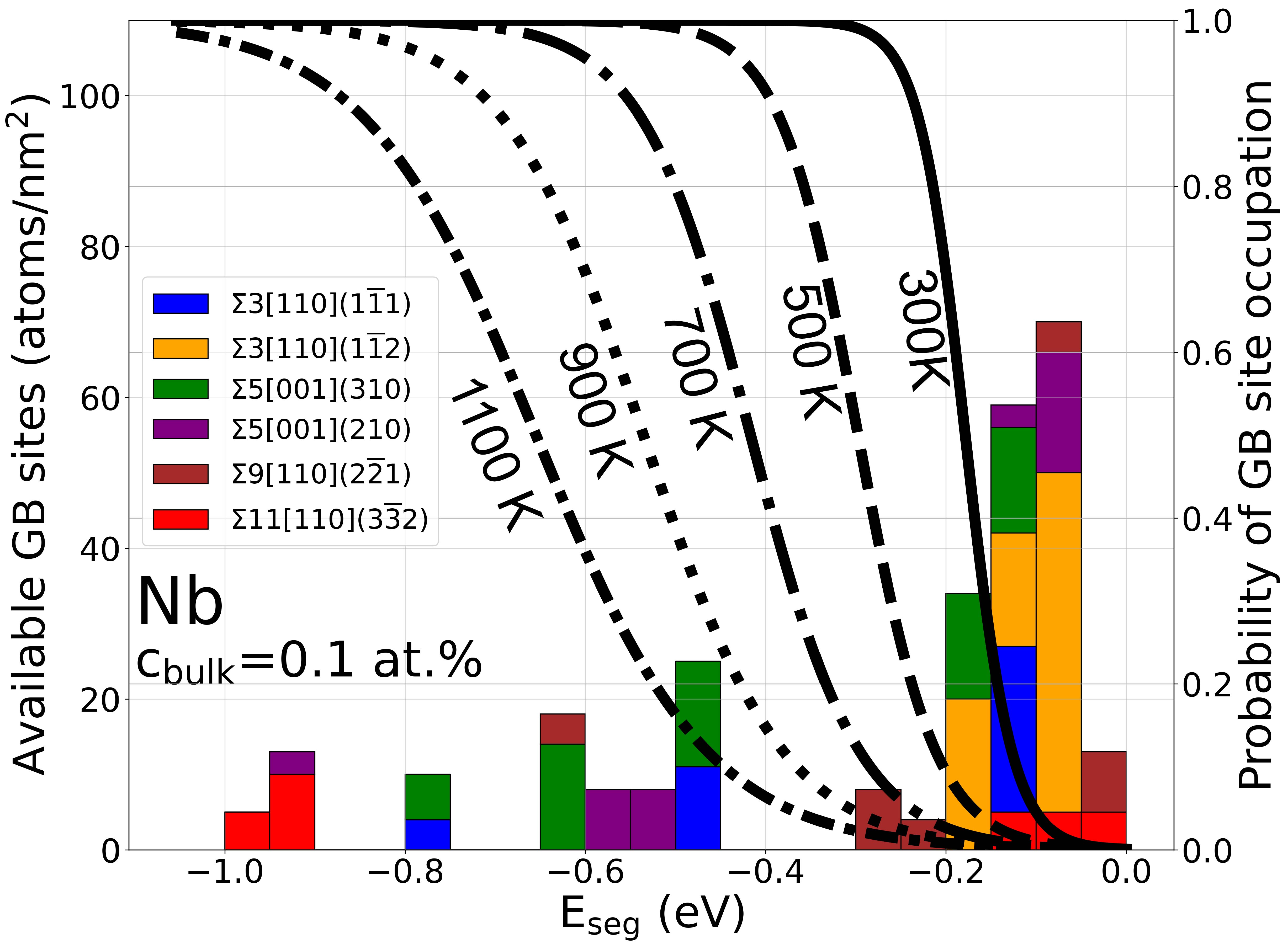}
		\caption{}
		\label{fig:SiteEsegDOS_LangmuirDistribution_Nb}
	\end{subfigure}
	\caption{Demonstrations of why scalar approaches to segregation modelling often fail to hold quantitative predictive power. The (\ref{fig:minEseg_vs_WhiteCoghlanAmount}) expected White-Coghlan multi-site isotherm interfacial coverage vs. the minimum segregation energy of elements at each GB in an Fe-X binary alloy containing 0.1 at.\% of the solute and the effective smeared segregation energy for (\ref{fig:TempEffectiveSegregationEnergy}) Nb and (\ref{fig:TempEffectiveSegregationEnergy_Mo}) Mo, that would be measured at varying temperatures, e.g., as in experiments utilising a Langmuir-McLean approximation. A reference line is plotted for the experimental value of 28, 38 kJ/mol measured at 800\degree C in ferritic iron at 0.087 at.\% Nb and 0.176 at.\% Mo respectively, in Ref. \cite{maruyamaInteractionSoluteNiobium2003}. (\ref{fig:SiteEsegDOS_LangmuirDistribution_Nb}) The density of states for the  available segregation sites for Nb is plotted against the probability of occupation, as calculated using the Langmuir-McLean isotherm at 0.1 at.\%.}
	\label{fig:TemperatureAdjSegregation}
\end{figure}
\clearpage
\subsection{Cohesion}

\subsection{Elemental effects}
\begin{figure}[h!]	
	\centering
	\makebox[\linewidth]{
		\begin{subfigure}{\linewidth}
			\includegraphics[width=1.0\linewidth]{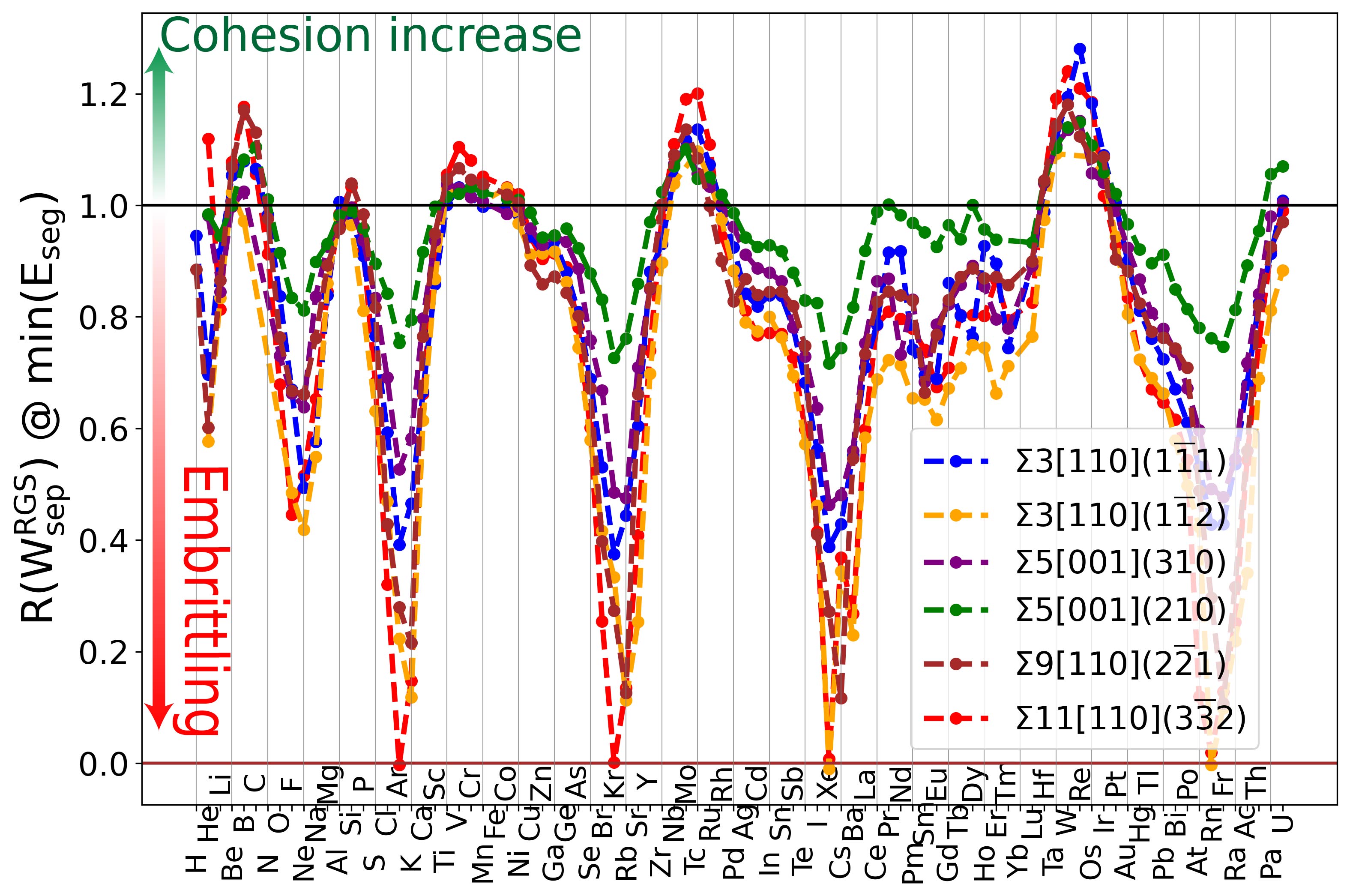}
			\caption{}
			\label{fig:mincohesion_rWsep_vs_Z}
		\end{subfigure}
	}
	\\ 
	\makebox[\linewidth]{
	\begin{subfigure}{\linewidth}
		\includegraphics[width=1.0\linewidth]{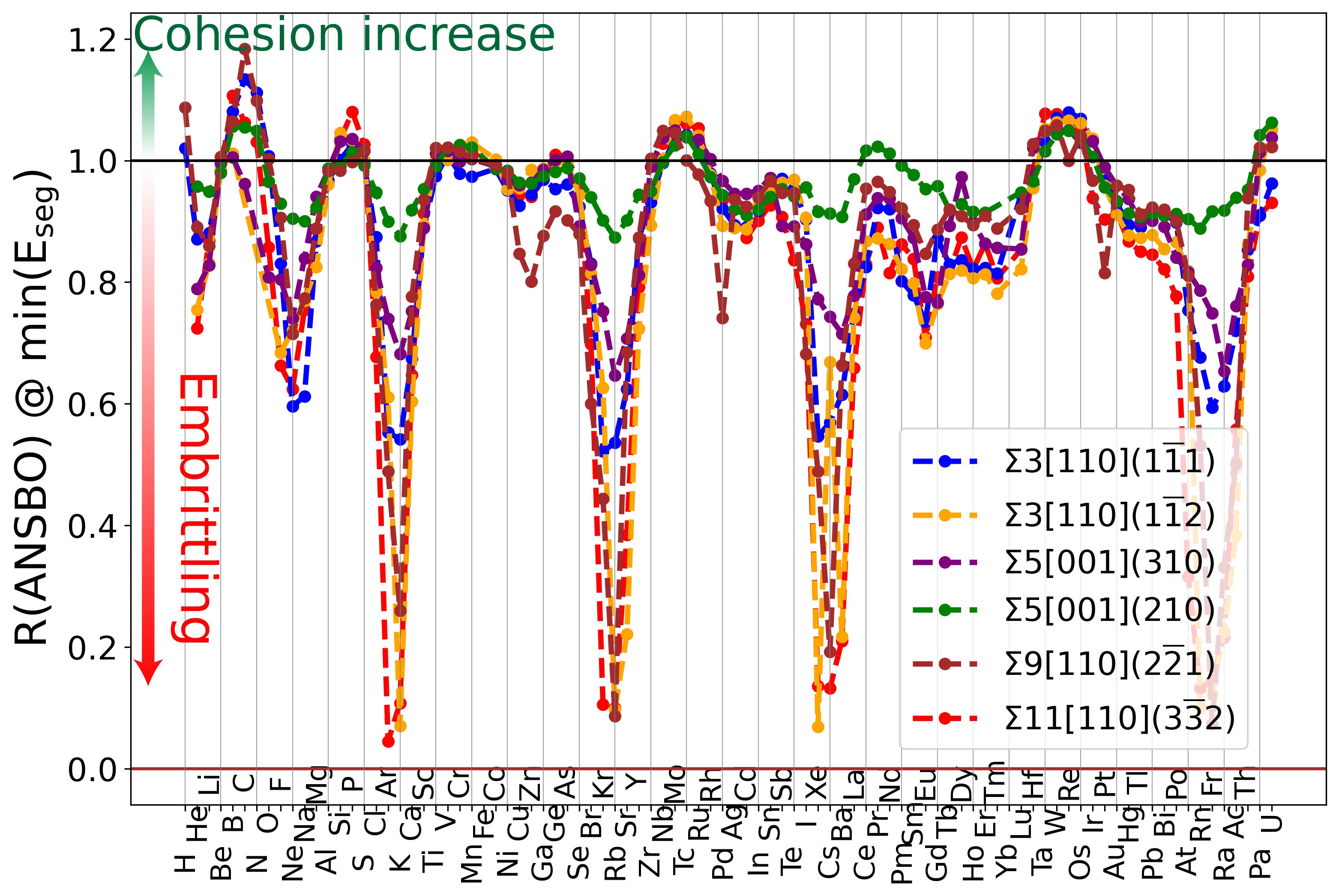}
		\caption{}
		\label{fig:mincohesion_rANSBO_vs_Z}
	\end{subfigure}
}
	\caption{The calculated elemental effects on GB interface cohesive strength, based on their strongest binding site at each GB, are plotted against the elemental number $Z$. The effect on cohesion is quantified by the relative cohesion to the pure GB (R), using (\ref{fig:mincohesion_rWsep_vs_Z}) the $\text{W}_{\rm sep}^{\rm RGS}$ framework and (\ref{fig:mincohesion_rANSBO_vs_Z}) the ANSBO framework. These values are available in Table~\ref{SI:tab:strongestbindingsites} in the Supplementary Information.
}
	\label{fig:mincohesion_vs_Z}
\end{figure}

The effect of each element on GB cohesion is of obvious engineering interest. One way of approaching this problem is to plot the effect that a solute has on interface cohesion at the site in which it is most strongly bound, as is done in Fig. \ref{fig:mincohesion_vs_Z}. Since only the \textit{lower} bound of interface cohesion is of engineering interest when solutes are segregated,  we plot only the smallest calculated cohesion quantity calculated over all separation planes tested for each site. The results for all other cleavage planes studied for each site are available in the Supplementary Data.

\subsection{Segregation engineering maps}
\begin{figure}[h!]
	\centering
	\begin{subfigure}{0.75\linewidth}
		\centering
		\includegraphics[width=\linewidth]{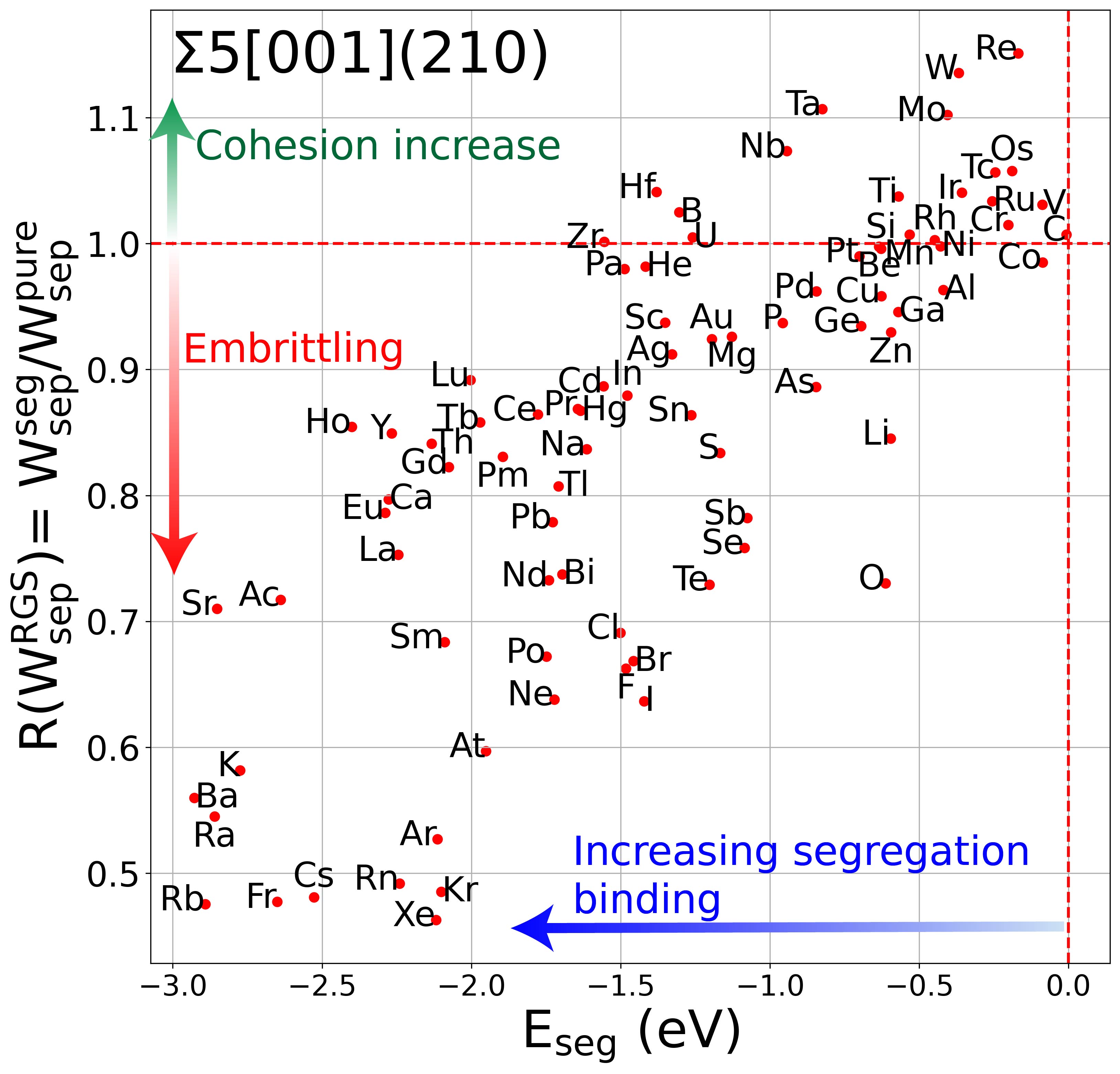}
		\caption{}
		\label{fig:RWsep_vs_Eseg_S5_210}
	\end{subfigure}
	\\
	\begin{subfigure}{0.75\linewidth}
		\centering
		\includegraphics[width=\linewidth]{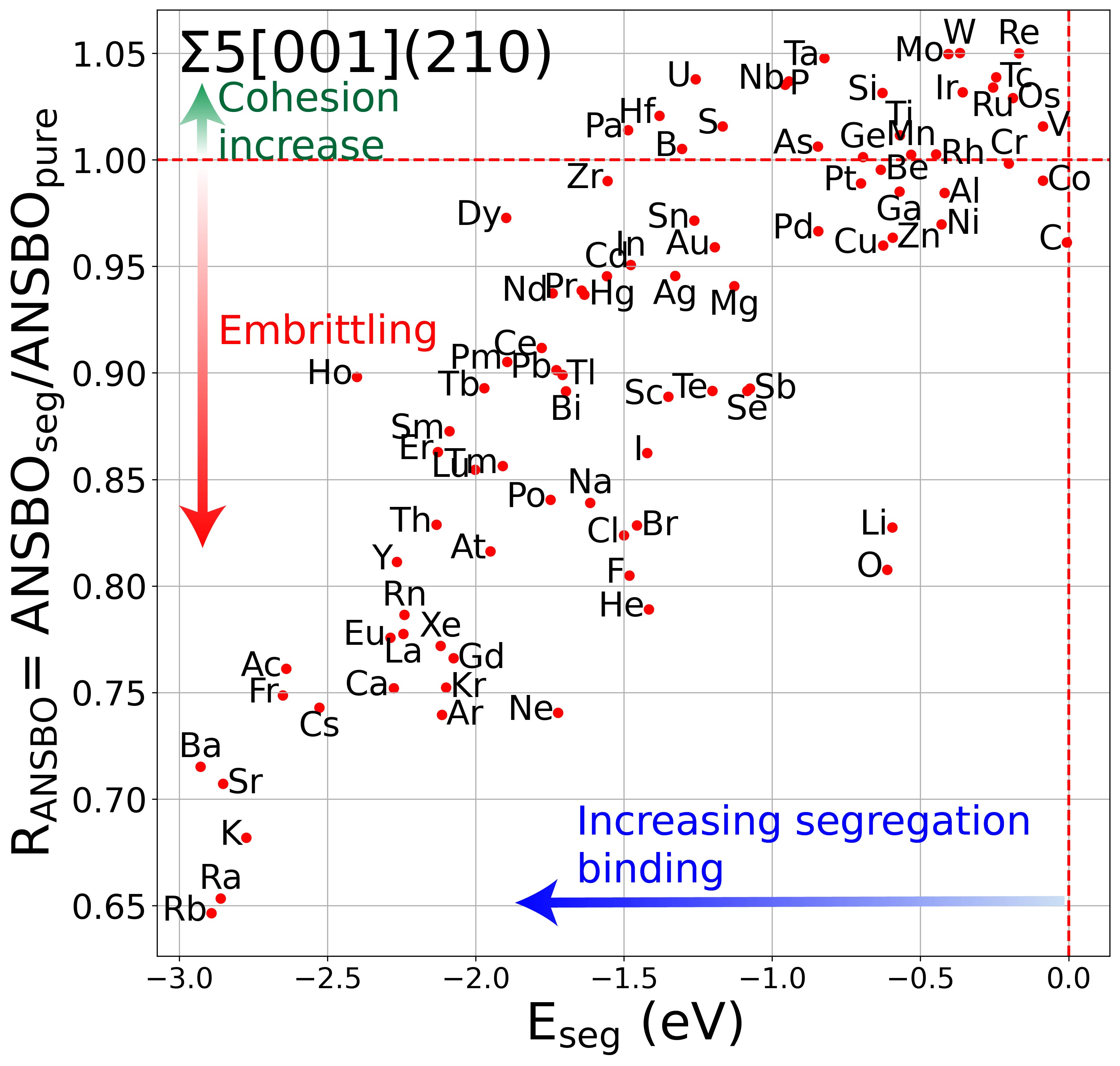}
		\caption{}
		\label{fig:RANSBO_vs_Eseg_S5_210}
	\end{subfigure}
	\caption{The enacted effects by solutes on GB cohesion at their strongest binding sites in the $\Sigma5[001](210)$ GB, quantified by the cohesion ratios between a segregated and a pure GB (\ref{fig:RWsep_vs_Eseg_S5_210}) the rigid Rice-Wang work of separation (W$_\text{sep}^\text{RGS}$) and (\ref{fig:RANSBO_vs_Eseg_S5_210}) the area-normalised DDEC6 summed bond orders (ANSBO), are plotted against their maximum segregation binding at a GB. The values on the y-axis are normalised with respect to the values in the pure GB, for direct comparison. Plots for each GB are available in Figs. \ref{SI:fig:RWsep_vs_Eseg_S3_111}, \ref{SI:fig:RANSBO_vs_Eseg_S11_332} in the S.I. Corresponding plots of the same data are available on a periodic-table visualisation for trend viewing in Figs. \ref{SI:fig:Ptable_segregation_engineering_ANSBO_bounded}, \ref{SI:fig:Ptable_segregation_engineering_Wsep_bounded} in the S.I. This data is available in tabulated form in the S.I.}
	\label{fig:cohesion_vs_Eseg_S5_210}
\end{figure}

\begin{figure}[h!]
	\centering
	\begin{subfigure}{0.75\linewidth}
		\centering
		\includegraphics[width=\linewidth]{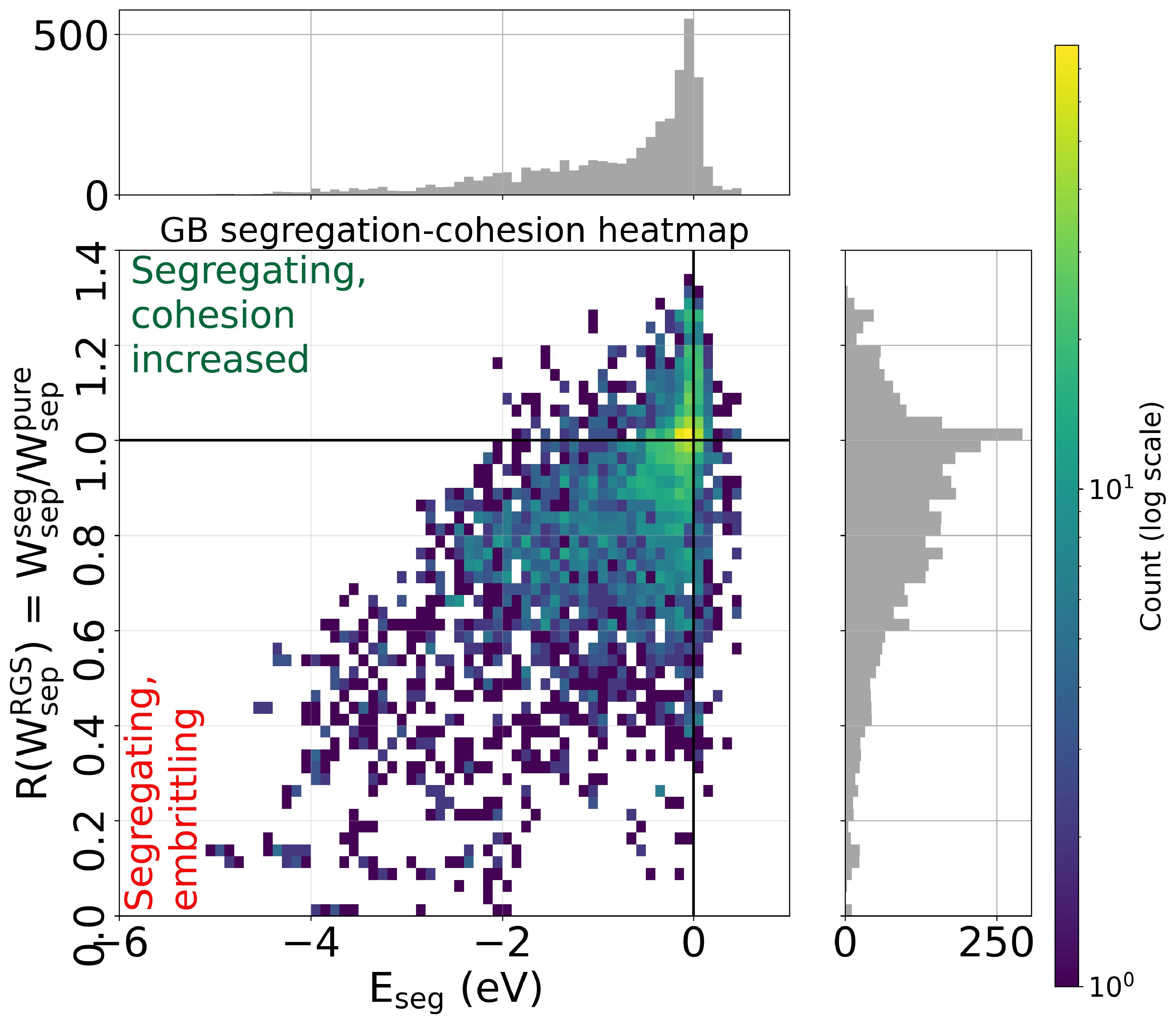}
		\caption{}
		\label{fig:RWsep_vs_Eseg_allsites}
	\end{subfigure}
	\\
	\begin{subfigure}{0.75\linewidth}
		\centering
		\includegraphics[width=\linewidth]{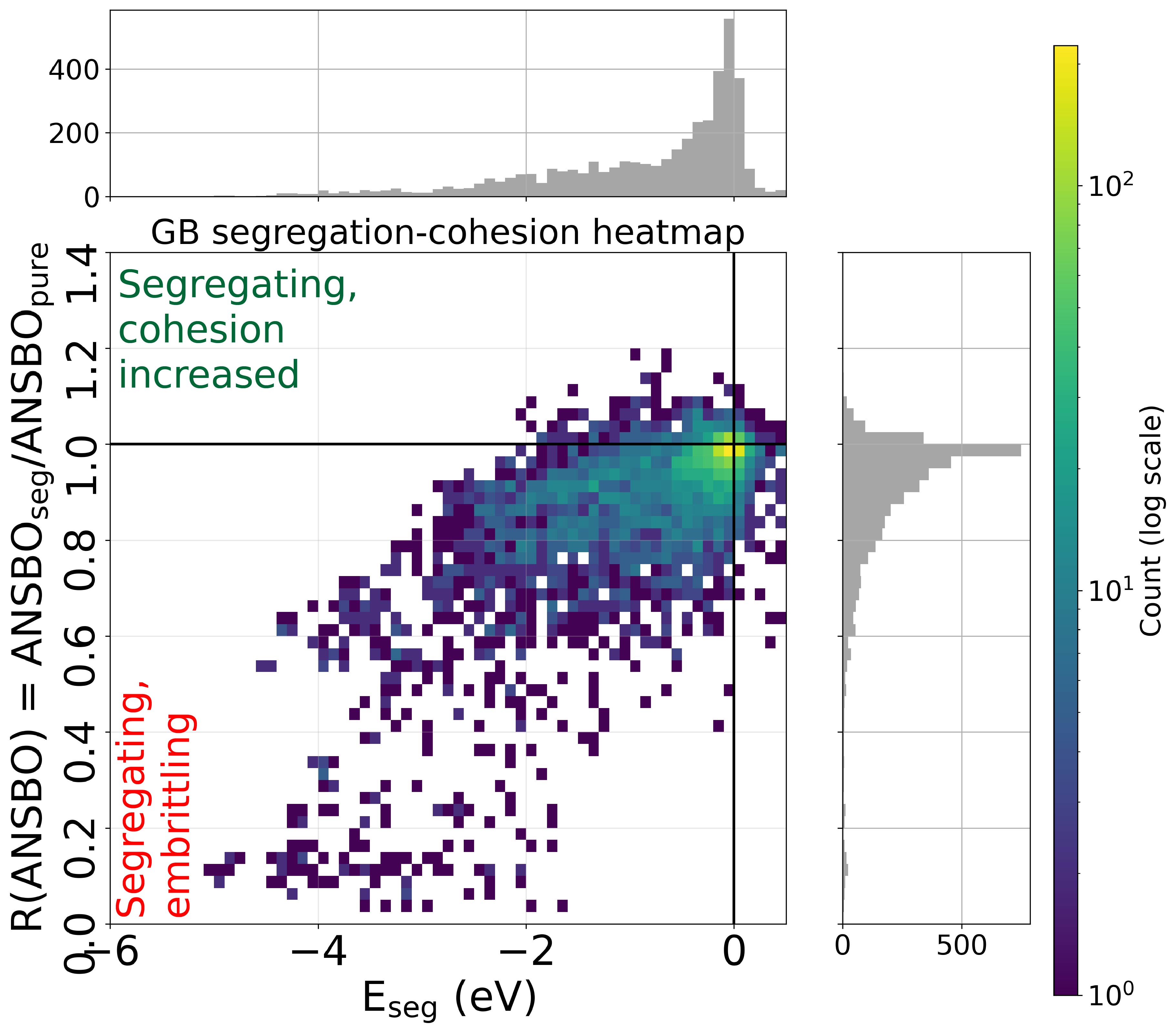}
		\caption{}
		\label{fig:RANSBO_vs_Eseg_allsites}
	\end{subfigure}
	\caption{The enacted effects by solutes on GB cohesion, quantified by the cohesion ratios (seg/pure) between a segregated and a pure GB (\ref{fig:RWsep_vs_Eseg_allsites}) the rigid work of separation (W$_\text{sep}^\text{RGS}$) and (\ref{fig:RANSBO_vs_Eseg_allsites}) the area-normalised DDEC6 summed bond orders (ANSBO), are plotted against their tendency to segregate E$_\text{seg}$ across all sites. The values on the y-axis are normalised with respect to the values in the pure GB, for direct comparison. Per-GB plots are available in Figs. \ref{SI:fig:RWsep_vs_Eseg_AllSites_Heatmap_S11_332}, \ref{SI:fig:ANSBO_vs_Eseg_AllSites_Heatmap_S11_332} in the S.I. This data is available in tabulated form (a python pandas DataFrame) in the Supplementary Data.}
	\label{fig:cohesion_vs_Eseg_allsites}
\end{figure}

\begin{figure}[h!]
	\centering
	\begin{subfigure}{\linewidth}
		\includegraphics[width=0.9\linewidth]{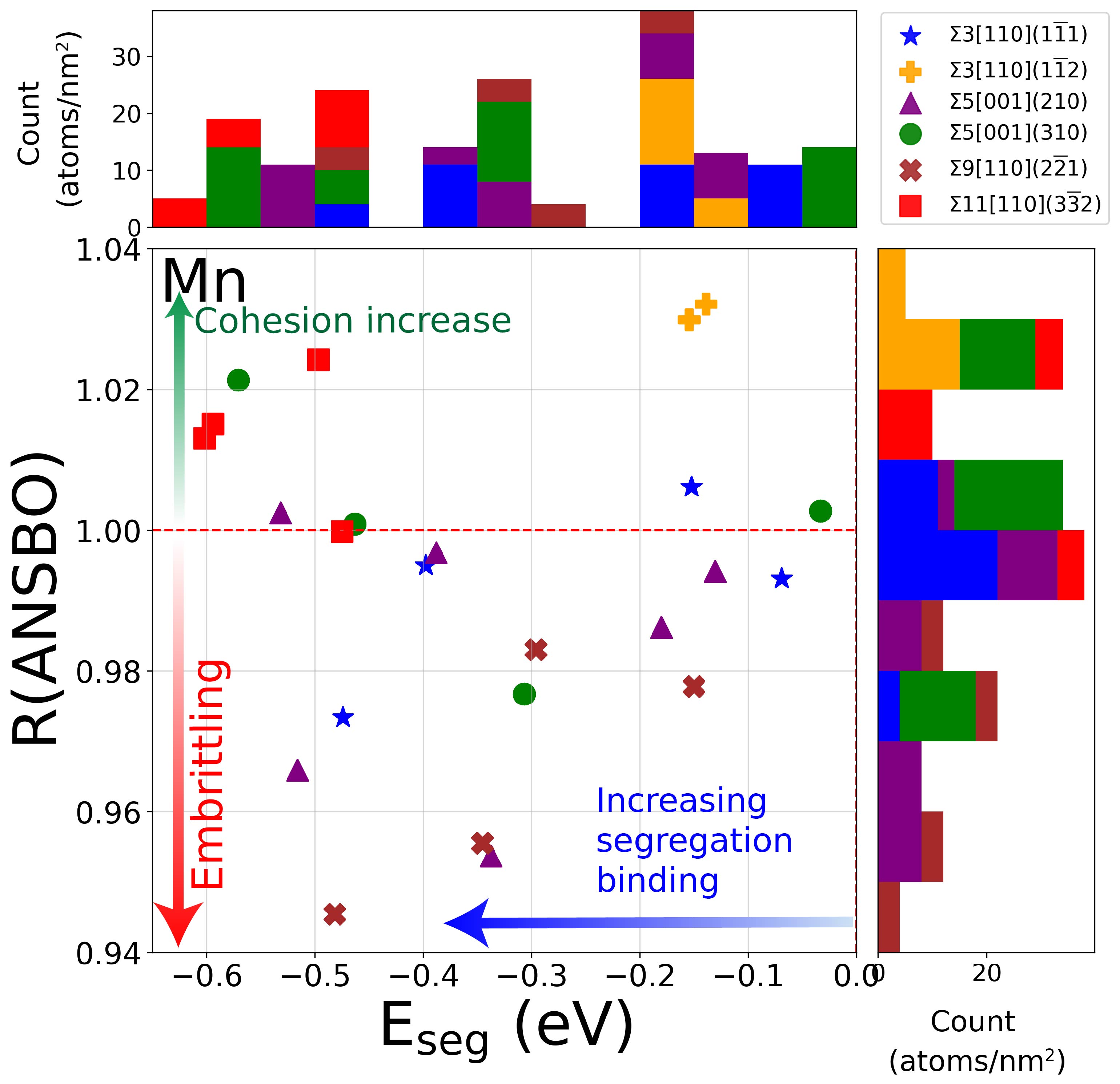}
		\caption{}
		\label{fig:SegregationEngineering_Mn_E_seg_vs_R_ANSBO}
	\end{subfigure}
	\\ 
	\begin{subfigure}{\linewidth}
		\includegraphics[width=0.9\linewidth]{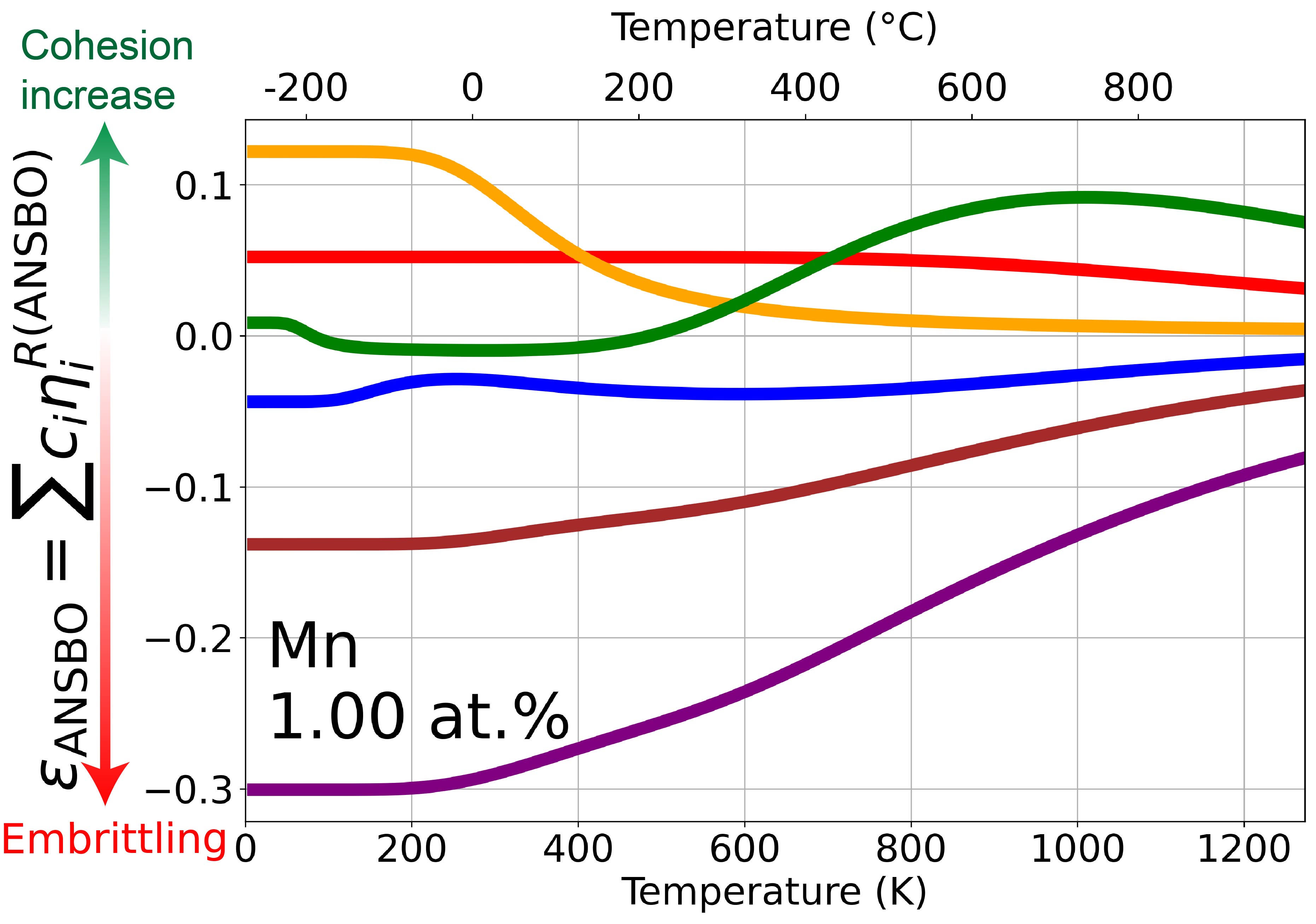}
		\caption{}
		\label{fig:TemperatureAdjCohesion_Mn_ANSBO}
	\end{subfigure}
	\caption{The (\ref{fig:SegregationEngineering_Mn_E_seg_vs_R_ANSBO}) segregation engineering map (segregation binding strength at a site vs. their effect on GB cohesion) for Mn at various sites and (\ref{fig:TemperatureAdjCohesion_Mn_ANSBO}) the temperature-adjusted induced cohesive effects, as evaluated in the DDEC6 ANSBO framework ($\epsilon_{\rm{ANSBO}}$) that segregated Mn enacts at 1 at.\% bulk concentration are plotted. The tabulated data, corresponding evaluation using the Rice-Wang \textit{rigid} work of separation framework and similar plots for all other elements can be made using the jupyter notebooks and data contained in the Supplementary Information.}
	\label{fig:TemperatureAdjCohesion_2x2}
\end{figure}

Of particular interest are the elements that exhibit strong segregation to GBs and enact significant beneficial or deleterious effects on interface cohesion. A subset of the data presented in Fig. \ref{fig:mincohesion_vs_Z}, specifically that pertaining to the $\Sigma5[001](210)$ GB, is presented in Fig. \ref{fig:cohesion_vs_Eseg_S5_210}. The effects that segregated solutes have on GB cohesion are plotted against their tendency to segregate across all sites in Fig. \ref{fig:cohesion_vs_Eseg_allsites}. From these plots, it is evident that the strength of segregation binding of a solute is generally correlated with a more detrimental effect on cohesion.
\\
However, just as a single segregation energy fails to capture how the distribution of sites alters the total amount of segregation [Fig. \ref{fig:WhiteCoghlan}]. Hence, considering only a single cohesive effect for a solute segregated in its most favourable position at a GB fails to capture its real effect on GB cohesion. A demonstration can be made by plotting the segregation energies of a single element, e.g. Mn, at various sites against their effect on cohesion [Fig. \ref{fig:SegregationEngineering_Mn_E_seg_vs_R_ANSBO}]. One can then utilise the cohesion and segregation data to calculate the thermodynamically weighted cohesive effects via equation \ref{eqn:TempAdjCohesion} in Methods, resulting in plots such as Fig. \ref{fig:TemperatureAdjCohesion_Mn_ANSBO} for Mn. Similar plots for other elements are available in the S.I.
\subsection{Cohesion in the ANSBO vs. Rice-Wang framework}
\begin{figure}[h!]
	\centering
	\begin{subfigure}{0.9\linewidth}
		\centering
		\includegraphics[width=\linewidth]{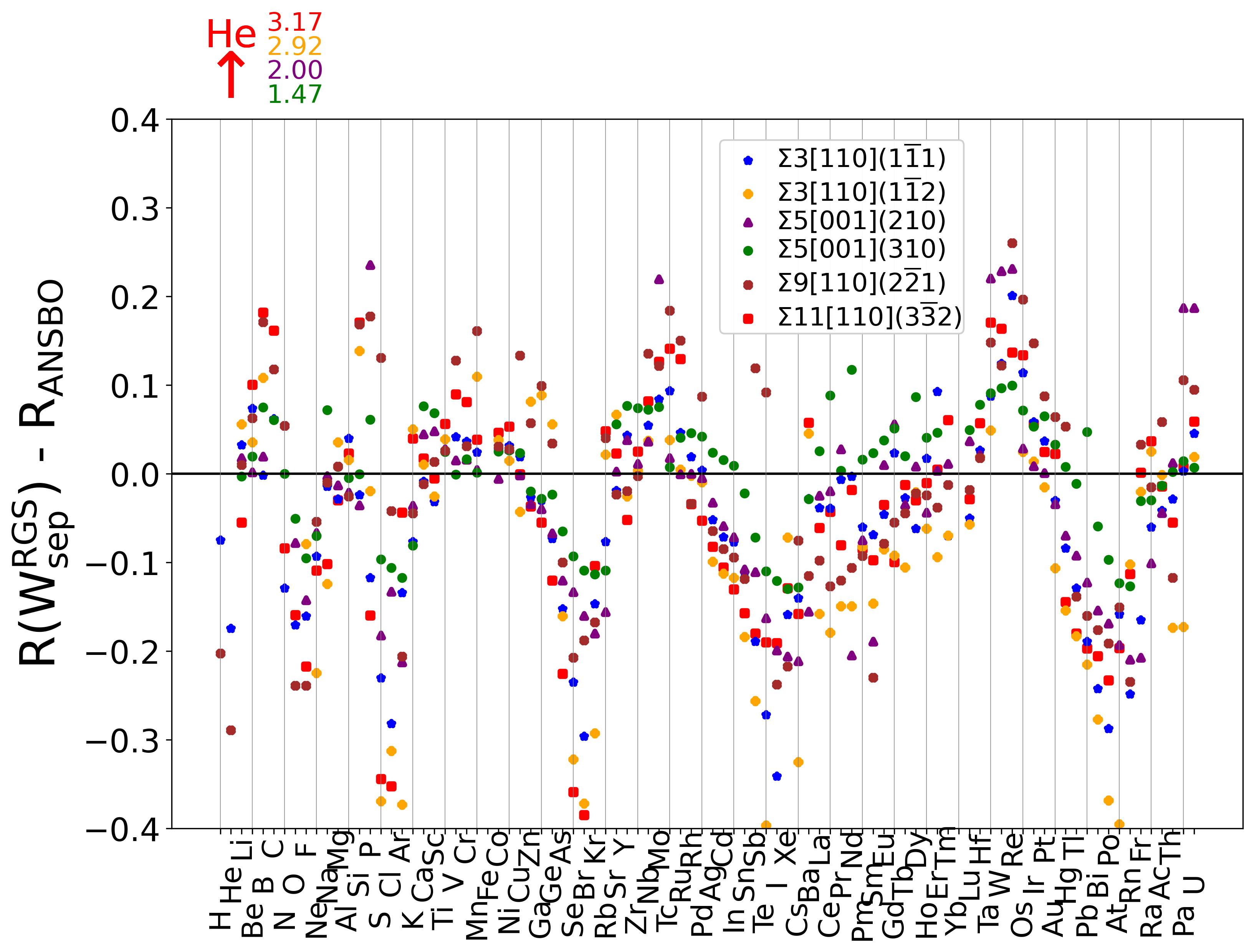}
		\caption{}
		\label{fig:err_Rwsep_R_ANSBO_vs_Z}
	\end{subfigure}
	\begin{subfigure}{0.9\linewidth}
			\centering
			\includegraphics[width=\linewidth]{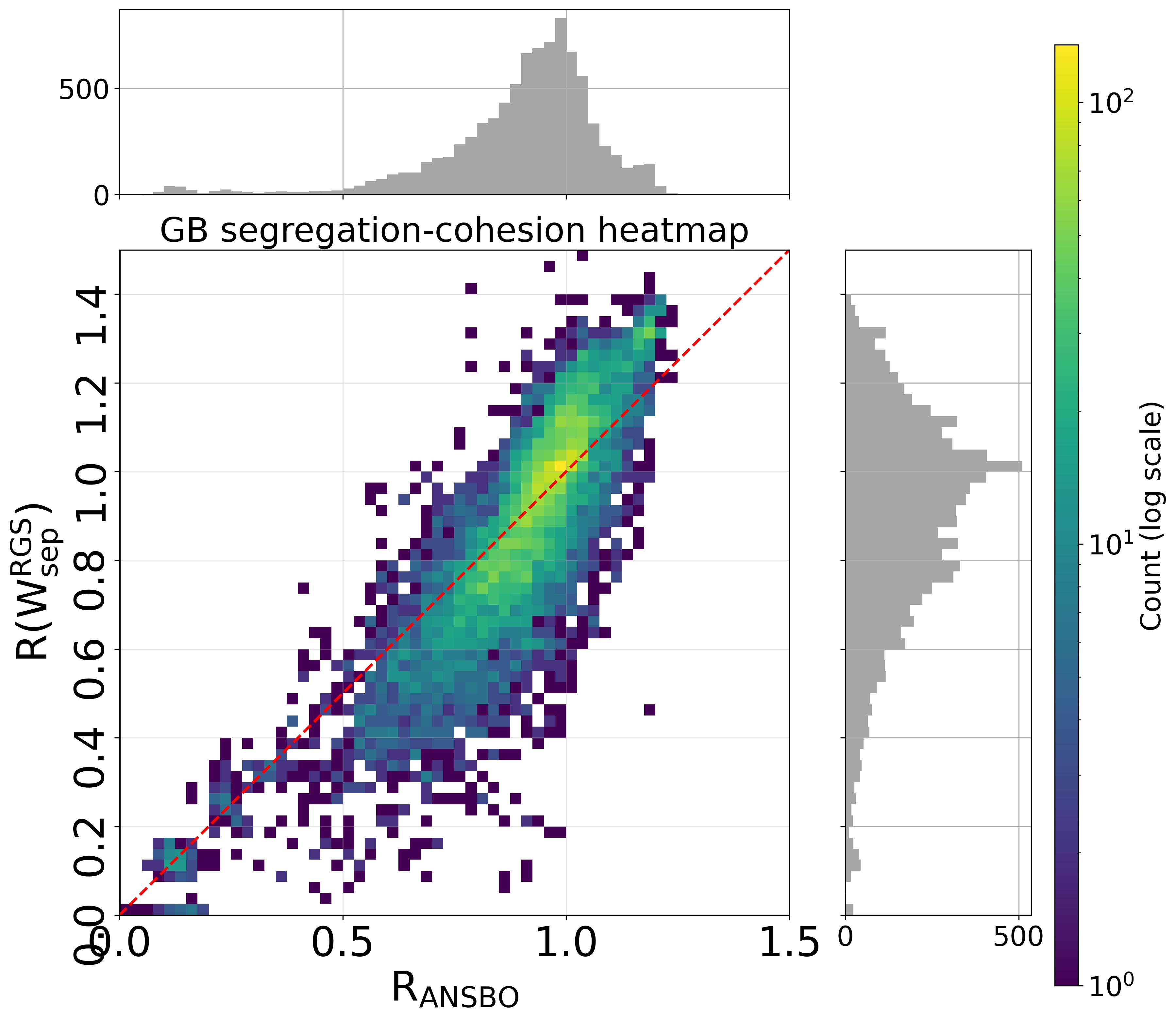}
			\caption{}
			\label{fig:R_Wsep_vs_R_ANSBO}
		\end{subfigure}
	\caption{Plots comparing the cohesion quantities calculated in the bonding-based DDEC6 area-normalised summed bond orders and the work of separation quantities calculated in the Rice-Wang framework. (\ref{fig:err_Rwsep_R_ANSBO_vs_Z}) The difference in the evaluated effects on cohesion, illustrated  at the strongest binding site in each GB, quantified by the differences in R$_\text{ANSBO}$ and R(W$_{\text{sep}}^\text{RGS}$), the ratios of cohesion between the segregated and pure GBs, are plotted against elemental number Z. (\ref{fig:R_Wsep_vs_R_ANSBO}) The 2D histogram showing the correlation between the quantities calculated at the three weakest cleavage planes (quantified by ANSBO) across all segregation sites with E$_{\rm{seg}} < -0.1$ eV. The values calculated at the coordinates of the weakest three cleavage planes (in the ANSBO basis) in both frameworks is used for comparison in each segregation case. Note that there are some outliers that lie outside the plot areas for both figures (e.g., He) which are explained in the text.}
	\label{fig:cohesion_comparison}
\end{figure}

The predicted effects on GB cohesion in each framework (Rice-Wang vs. ANSBO) at each maximum segregation binding site studied for all elements are compared in Fig. \ref{fig:err_Rwsep_R_ANSBO_vs_Z}. The cohesion values of the three smallest, evaluated by $\text{R}_{\text{ANSBO}}$, and the corresponding $\text{R}_{\text{W}_\text{sep}^\text{RGS}}$ at those planes are plotted against each other across all studied segregating cases (E$_\text{seg}$ $<$ -0.1 eV) in Fig. \ref{fig:R_Wsep_vs_R_ANSBO}. Here, we look at only the planes which are predicted to be the weakest via DDEC6 theory, and the corresponding W$_{\text{sep}}$ value at the same coordinate plane, i.e. \textit{not} the weakest planes in both frameworks. This distinction is important, since in 98\% of cases, the predicted weakest cleavage planes are different in the two frameworks. The purpose of such an analysis is to evaluate the agreement or disagreement in the predicted elemental cohesive effects at GBs in both theoretical frameworks. 
\\\\
\clearpage
\section{Discussion}
\subsection{Insights into segregation}
As a rule of thumb, the maximum strength of segregation binding at GBs trends higher with increasing solute size compared to the host bulk Fe [Figs. \ref{fig:Eseg_vs_VorVolBulk}, \ref{fig:Ptable_corrcoefs_EsegVorvol}]. In the case of the special twin-like GBs which contain only bulk-like sites, this relationship does not hold, and very minimal segregation trapping overall is observed in these GBs. We speculate that the maximum segregation binding in other metallic host lattices can also be roughly estimated in this manner. Elements with the strongest segregation bindings are generally the ones that are extremely large compared to Fe, and thus energetically highly unfavourable when placed in the bulk lattice.
\\\\ 
Significantly, we find that the Voronoi volume of sites can be used to qualitatively estimate overall segregation binding tendencies at \textit{different} sites in GBs for many elements across the periodic table [Fig. \ref{fig:Ptable_corrcoefs_EsegVorvol}]. The large (both positive and negative) rank and linear correlation coefficients across significant parts of the periodic table indicate that relative differences in substitutional segregation binding for the same element can be described via a simple linear site-volume argument at the site occupied by the solute. Although statements to this effect have been made previously \cite{jinStudyInteractionSolutes2014, yangHighthroughputFirstprinciplesInvestigation2023}, we demonstrate comprehensively here for which elements this relationship functions, and to what extent, with around 100 datapoints for each element. 
\\\\
Due to the often giant relaxations triggered by many segregating elements at GBs, the Voronoi volume relationship is only valid for relaxed structures and not for predicting segregation energies \textit{a-priori} from the pure GB structures. Nevertheless, this fact may be exploited to predict both the strength of segregation binding of many elements and their relative differences at differing sites in experiments that are able to measure the volume of solute-occupied sites at GBs with atomic level microscopic resolution. For example, such volumes can be accessible via TEM images, which can differentiate the volumes between sites, and hence, likely the segregation energies \cite{ahmadianAluminumDepletionInduced2021}. This can provide a relatively accurate qualitative picture of \textit{site}-level segregation strengths for these elements without conducting further experiments to analyse segregation binding strength. Since single site-segregation energies, and hence segregation spectra, are currently inaccessible experimentally, with most techniques accessing area-averaged concentrations instead, this should serve as a valuable guide for experimentalists overall.
\\\\
Importantly, the trends of segregation coverage are not well predicted for most of the periodic table when only looking at trends for maximum segregation binding, as is often presented in the literature. A good example would be the differences in the maximum segregation binding of the transition metals and the lanthanides. Taking a naive approach, by looking at only their maximum segregation energies [Fig. \ref{fig:minEseg_vs_Z}], one would expect at least double the amount of segregated lanthanides compared to the transition metals in their respective Fe-X binary alloys. However, a closer look at their expected segregation coverages, calculated via the White-Coghlan isotherm at a hypothetical scenario of Fe-X 0.1 at.\%\ alloy at 300K reveals that the coverages are commonly within 10 atoms/nm$^2$ of one another in certain GBs [see Supplementary Fig. \ref{SI:fig:WhiteCoghlan_vs_Z_300K_0.1at}]. 
\\\\
To understand the weak correlation between the strength of maximum segregation binding and total segregation amount at a GB [Fig. \ref{fig:minEseg_vs_WhiteCoghlanAmount}], there are two important facts to consider; the relevance of bindings above 0.8~eV in strength, and the spectrum of available sites to bind solutes at a GB. Firstly, segregation bindings of E$_\text{seg} < -0.8$ eV are generally irrelevant in site occupation considerations as a function of temperature [see Supplementary Fig. \ref{SI:fig:Mclean_vs_Eseg}]. This is because a E$_\text{seg} = -0.8$ eV at a site already guarantees full occupation at 0.1 at.\% at temperatures of almost 700\degree C in binary systems. These temperatures are generally at the upper bound of operating temperatures of ferritic steels. Stronger bindings are therefore only relevant in discussions around the site-competition effects between different solutes, and GB stabilisation considerations, e.g., in nanocrystalline alloy design criteria. Secondly, it is important to recognise that solutes exhibiting moderate segregation trapping across more sites, can result in comparable total segregation coverages to those that exhibit extremely large segregation trapping at fewer sites at a GB. In the extreme cases of the extremely large solutes, such as those belonging to the actinides or noble gases, they can observe very deep segregation trapping at GBs but only at a few sites. In these cases, the overall coverage at the GB for these elements may still be comparable or lower than that of elements with weaker maximum binding at the GB, but bind at more sites. This assertion is valid as long as there is enough solute in the alloying composition to saturate the deepest traps. Since the specific statistics on the  site distributions in a polycrystal are not the focus of the study here, we do not discuss when this may occur.
\\\\
Here we should note that solute-solute interactions are mostly repulsive in nature \cite{maiSegregationTransitionMetals2022, maiPhosphorusTransitionMetal2023}, and are largest for larger solutes \cite{maiSegregationTransitionMetals2022}. As such, we speculate that the explicit inclusion of such solute-solute interactions, will mostly result in an exacerbation of this specific phenomena. A general rule of thumb would be therefore to consider models excluding such solute-solute interactions to generally predict an upper bound of segregation coverage that may be observed at GBs. Specific situations considering precipitation or cases where high-concentrations of solutes are present at GBs, where solute-solute interactions can play an outsized-role, need to be considered explicitly on a case-by-case basis, which we leave for future work. 
\\\\
We now explain, using examples from the literature and Figs. \ref{fig:SiteEsegDOS_LangmuirDistribution_Nb}, \ref{fig:TempEffectiveSegregationEnergy}, why the practice of assigning single segregation energies for solutes at GBs often results in significant discrepancies in both experiment and simulations. Experiments often prescribe segregation energies via the Langmuir-McLean equation, assuming only a single effective site at a GB [e.g., \cite{maruyamaInteractionSoluteNiobium2003}]. However, this neglect of the spectrum of binding energies that exists at GBs therefore results in an artificial temperature dependence of the measured segregation \textit{enthalpies} in experiments, which cannot be isolated from the true entropic contributions [e.g., as is seen in the effective segregation energy in Fig. \ref{fig:TempEffectiveSegregationEnergy}]. On the other hand, simulation-based studies of segregation also occasionally only parameterise segregation with scalar approaches, i.e., presenting single values of segregation at a specific site \cite{huSoluteEffectsS32020, pengCorrelationStabilizingStrengthening2023}, or at their strongest binding site. However, weaker sites must play a significant role in the observed solute coverage at GBs, as is seen in the density-of-states/site-occupation probabilities charted for the example of Nb in Fig. \ref{fig:SiteEsegDOS_LangmuirDistribution_Nb}. This practice then often results in an apparent severe overestimation of segregation binding when comparing DFT computed energies with those extracted or derived from experiments, such as by Auger electron spectroscopy \cite{lejcekWhyCalculatedEnergies2013, lejcekInterfacialSegregationGrain2017} or atom probe tomography (APT) \cite{maruyamaInteractionSoluteNiobium2003}. 
\\\\
The density of states of Nb binding sites in our study is presented in Fig. \ref{fig:SiteEsegDOS_LangmuirDistribution_Nb}, plotted against the Langmuir-McLean isotherm predicted probability of occupation. The maximum segregation binding of Nb at GBs of $0.7-0.9$~eV [Fig. \ref{fig:minEseg_vs_Z}] appears to be significantly stronger when compared to the experimentally derived values of $0.37-0.41$~eV in atom probe experiments \cite{maruyamaInteractionSoluteNiobium2003} containing 0.087 at.\% at 800\degree C. The Langmuir-McLean isotherm was employed in order to obtain these segregation energies, assuming a single type of site, with no solute-solute interactions, calculating the Gibbs free energy of segregation with the observed interfacial excess of the solute from APT. However, the effective segregation energy (i.e., enthalpy only) accounting for the entire spectra of binding sites is generally on the order of about half that, at $0.45-0.55$~eV, when accounting for the spectral nature of segregation [Fig. \ref{fig:TempEffectiveSegregationEnergy}], which is then in much better agreement with the \textit{free energies} in the experiments, even without accounting for other finite temperature contributions (paramagnetism, etc.). A similar comparison may be made for Mo at 0.176 at.\% at 800\degree C, with derived segregation energies of $-0.26 - -0.31$~eV from experiments \cite{maruyamaInteractionSoluteNiobium2003}, and $-0.4- -0.5$~eV from DFT. In this case, similar good agreement is achieved when looking at the \textit{effective} segregation energies of $-0.2--0.3$~eV when accounting for spectra of binding sites available at GBs. Thus, the apparent overestimation from DFT when compared to experiment can simply be a failure to account for the fundamentally spectral nature of segregation, due to the common practice of using the \textit{maximum} segregation binding from DFT studies for comparison, instead of comparisons which account for their \textit{spectra}. We expect this argument to hold similarly for interstitial solutes as well, where the apparent overestimation in binding is most severe \cite{lejcekWhyCalculatedEnergies2013, lejcekInterfacialSegregationGrain2017}.
\\\\
Our study reveals that in chemically complex alloys, it is not only the \textit{spectrality} of segregation binding energies that is important in modelling segregation phenomena, but also their preferred occupation sites. Since smaller solutes (e.g., B, C, P and S ) and larger solutes (e.g., Mo, Ti, W ...) prefer inherently different types of sites, site competition effects in complex alloys cannot be simply described by a simple spectrum distribution model. Small solutes tend to occupy small sites and often exhibit anti-segregation behaviour to larger sites, whereas the inverse is true for larger solutes. There are important smaller solutes which include powerful embrittlers (e.g., O, He) and strengtheners (e.g., B, C) from the light-elements, which occupy different types of sites compared to many typical alloying elements. Therefore, any segregation engineering attempts aiming to prevent the enrichment of embrittling solutes at GBs need to additionally consider the discrete nature of site environments and solute site preferences. A simple and natural extension of existing segregation isotherms for multi-component alloys would be to partition the two kinds of sites for different types of elements (e.g., small vs large). However, it should be noted that there are a few solutes (e.g., Ni, Co, Mn, Cr, Si, V) where site-preference is not dominated by size, and therefore require more complex model features, such as those involving magnetism and chemical bonding.
\subsection{Effects of segregants on cohesion}
From Figure \ref{fig:cohesion_vs_Eseg_allsites}, what is immediately apparent is that generally elements with exhibit significant segregation tendencies also tend to be strong embrittlers. In fact, almost all elements which segregate strongly to GBs tend to be neutral to embrittling in nature. Thus, the focal point for alloy engineering should generally be the prevention of embrittling segregants from appearing at all at GBs. This can be achieved by trapping such elements in other phases, preventing their inclusion in alloys during synthesis, or employing site-competition-based segregation engineering, by including cohesion-neutral solutes with have stronger bindings to the same preferred sites. Critically, our results demonstrate that the susceptibility to segregation-induced embrittlement and strengthening varies significantly across different GBs. In essence, solutes which are embrittling in certain GBs can act neutral or strengthen in others.
\\\\
Of paramount importance is understanding the intrinsic complexity involved with segregation-induced cohesive changes at GBs. We find that segregation at different sites intrinsically results in variance in the expected cohesive effects, in both frameworks, of an element on GBs [see S.I. Figs. \ref{SI:fig:Eseg_vs_cohesion_ANSBO_H_vs_Eseg}-\ref{SI:fig:Eseg_vs_cohesion_ANSBO_U_vs_Eseg}, \ref{SI:fig:Eseg_vs_cohesion_wsep_H_vs_Eseg}-\ref{SI:fig:Eseg_vs_cohesion_wsep_U_vs_Eseg}]. Furthermore, due to the spectral nature of segregation binding, these sites are occupied at different concentrations at different alloying compositions and temperature. Even for the same element, the nature of its cohesive effect can range from embrittling to strengthening at different sites at GBs. Moreover, sites with similar segregation energies can possess markedly different effects on GB cohesion when hosting segregants [e.g., see Mn in Fig. \ref{fig:SegregationEngineering_Mn_E_seg_vs_R_ANSBO}]. It follows that generic or sweeping statements that effectively reduce the effects of solutes on GB cohesion to scalar quantities can only be valid in the specific conditions of their specimens, samples, or simulations. To accurately capture the effects of elements on GB cohesion, one must account for both the spectra of segregation binding \textit{and} enacted cohesive effects in tandem.
\\\\
Many elements may exhibit both embrittling and strengthening behaviours, depending on the specific sites that they are segregated to. By extension, the local structure of GBs, which govern the distribution of such sites, can alter the cohesion response of GBs to segregation. This contrast may be seen most prominently in the $\Sigma5[001](310)$ GB, which is clearly more resistant to cohesion reduction than the other GBs, even ones that are similar in GB energy, e.g. $\Sigma5[001](210)$ [Fig. \ref{fig:mincohesion_vs_Z}]. Since fracture is dependent on inherently localised atomistic features, mean-field approaches are unlikely to yield precise results with respect to solutes effects on cohesion. The exception is in the case where a specific element exhibits solely monotonous strengthening or weakening behaviours and segregation binding behaviours, which requires extensive computational surveying to confirm. Therefore, the confusion and scatter in the literature may be partially rationalised by the intrinsically complex nature of GB cohesion's response to segregation, as opposed to solely the unreliability of data.
\\\\
As an example, considerable confusion exists in the literature on whether Mn embrittles GBs \cite{kuzminaGrainBoundarySegregation2015,grabkeEffectsManganeseGrain1987, gelinasCorrelationFractureMechanisms2018}. In Fig. \ref{fig:TemperatureAdjCohesion_2x2}, we have shown that Mn's effect on cohesion varies from strengthening to embrittling, dependent on the type of GB that it is segregated to. Furthermore, even in the same GB, Mn can exhibit neutral, embrittling, or cohesion enhancing characteristics depending on the temperature [Fig. \ref{fig:TemperatureAdjCohesion_Mn_ANSBO}]. This is because its occupation at different sites, which may have differing effects on cohesion, changes depending on the temperature (e.g., heat treatment). By considering the ensemble of the cohesive effects at these sites as a thermodynamic average, it is demonstrated that Mn can vary from strengthening to neutral, even in the same GB, as shown in the $\Sigma5[001](310)$. Naturally, this would lead to different interpretations of whether it is an embrittling or neutral element for GB cohesion, even in otherwise tightly controlled experiments. For example, the $\Sigma5[001](310)$ and $\Sigma5[001](210)$ GBs, even though they differ in only 0.07 J/m$^2$ in GB energies, can display markedly different responses to Mn segregation. 
\\\\
Of alloy engineering interest are elements which exhibits both segregation and strengthening at GBs. Interestingly, such elements exist mostly at the refractory block of the transition metals, with Nb, Mo, Ta, W, Re and Os being generally beneficial to cohesion. Notably, these elements also tend to exhibit the lowest solute-size differential in the Fe-bulk in their respective rows in the periodic table [Fig. \ref{fig:VorVolBulk}]. The most detrimental elements for cohesion are those belonging to the alkali metals, alkaline earth metals, and the noble gases. This assessment of alkali metals agrees with the thermodynamic-model based predictions made by Gibson and Schuh \cite{gibsonSegregationinducedChangesGrain2015}. We find that their potency as embrittlers can quite simply be attributed to a size-effect; their presence at a GB forces neighbouring Fe-Fe bonds that are normal to the GB (and hence play a role in cohesion) to become considerably weaker (stretched) due to the large size of the atoms, confirmed in our ANSBO analysis. This applies to the alkaline earth and noble gas elements as well. 
\\\\
Of particular practical interest in the context of scrap recycling, are the so-called tramp elements P, S, Sn, Sb and As. The segregation binding of these elements is particularly strong compared to commonly used alloying transition metals, and they can therefore can be expected to out-compete the others at GBs when included in recycled iron alloys. The embrittling natures of Sn, Sb and As are particularly problematic because of this. Notably, S and P, two common suspects in temper embrittlement, are not predicted to have strong embrittlement effects here. However, at least in the case of P, we have previously demonstrated that more complex scenarios involving second-order effects, such as the repulsive nature of P-transition metal solute-solute interactions at GBs, need to be considered to elucidate their embrittling effect on GB cohesion in steels \cite{maiPhosphorusTransitionMetal2023}. As such, despite providing a thorough evaluation across binary Fe-X chemical space, solutes which embrittle GBs through such second-order effects may be overlooked by the results of our investigation here.
\\\\
Lanthanides (rare earths) tend to be mostly embrittling; however, in certain cases they may be neutral or even slight cohesion enhancers [e.g., Ce, Pr, Nd in the $\Sigma5[001](310)$]. Of the lanthanides, those with the smallest size tend to be the least detrimental to cohesion; whereas the largest tend to be the most detrimental. Furthermore; the strength of their segregation binding at GBs tends to be larger than all but the group 1, and 2 metals, as well as the noble gases. However, considerable differences in the strength of segregation binding exist amongst the lanthanides, which may be contrary to conventional belief in their similarity. Due to their strong segregation binding, they can be expected to be present at GBs whenever they are introduced into ferritic steels, if not captured by other phenomena, such as oxide formation, and generally be expected to embrittle them, either through a direct embrittlement effect, or through site-competition, by displacing cohesion enhancing/neutral transition metal solutes.
\\\\
The two theoretical frameworks for calculating cohesion may display surface-level qualitative agreement on cross-element trends across the periodic table [Fig. \ref{fig:mincohesion_vs_Z}], but significant discrepancies also emerge [Fig. \ref{fig:cohesion_comparison}]. In general, the cohesion enhancing effect for elements is generally predicted to be stronger in the rigid Rice-Wang framework than compared to the corresponding effect predicted by DDEC6 bond-orders. However, elements which have negative effects on cohesion are generally predicted to be more harmful in the DDEC6 framework compared to the rigid Rice-Wang [Fig. \ref{fig:R_Wsep_vs_R_ANSBO}]. So in general, rigid Rice-Wang has a bias towards larger cohesion values at interfaces compared to the DDEC6 framework, which comparatively tends to bias towards smaller values of cohesion effects.
\\\\
Let us pay special attention to the case of He, where the ease of the interpretation of the physics and chemistry in this case allows further physical analysis of the theories. When He is segregated, it does not form any significant bonds with the neighbouring Fe atoms, which is confirmed by DDEC6 bonding analysis, as is shown in Fig. \ref{SI:fig:He_cohesion_prediction_failure} in the S.I. The bonding-based DDEC6 theory therefore predicts \textit{embrittlement} at the GB as induced by He, as one would expect. However, unreasonably high surface energies, caused by our rigid-surface model, can cause unexpected predictions in the Rice-Wang framework [Fig. \ref{fig:err_Rwsep_R_ANSBO_vs_Z}]. Indeed, in certain cases, this high surface energy may lead to an unphysical prediction of \textit{strengthening} at a GB that may occur to He segregation. A similar situation also occurs in some instances with some other noble gas elements. The absence of bonding must lead to \textit{easier} brittle fracture along the GB, not harder. 
\\\\
This contradiction highlights a rather fundamental fact in that the difference in the surface and GB segregation energies need not necessarily be related to the effect that segregation has on fracture processes, which is concerned with \textit{bond-breaking}, as has been pointed out before \cite{mcmahonTheoryEmbrittlementSteels1978, briantChemistryGrainBoundary1990}. Note that we have previously demonstrated that using rigid surfaces to compute the Rice-Wang work of separation are in better agreement with bond-cleaving arguments than that using relaxed surfaces in various systems \cite{maiSegregationTransitionMetals2022, maiPhosphorusTransitionMetal2023}. However, this approach can also yield unreasonable results for He, and other noble gases, due to the high surface energies that are produced. Relaxing the surfaces, e.g., allowing release of the noble gas elements from the surface, must lead to a more realistic embrittlement factor from He computed from Rice-Wang. However, to treat fracture reversibly, the work to fracture must be to produce non-equilibrium surfaces \cite{mcmahonTheoryEmbrittlementSteels1978}. Then, the choice of which unrelaxed surface model must be used in Rice-Wang theory, is largely up to the discretion of the scientist. Consequently, this implies that the judgement on which surface model should be used in Rice-Wang calculations must be made on a case-by-case basis, as certain assumptions are more appropriate for specific chemical systems and situations.
\\\\
While these assumptions must be made and tested explicitly in Rice-Wang theory, here we highlight the fact that no such assumptions are required to capture the expected GB cohesion effects (e.g., the noble gases) in the quantum-chemistry bond-order based approach to quantifying effects on cohesion via the DDEC6 ANSBO cohesion theory. Moreover, we have found that the electronic convergence of certain surface calculations can be extraordinarily difficult for certain types of systems (e.g., noble gases, lanthanides). These DDEC6 calculations are computationally significantly cheaper, as only a single static DFT calculation to derive the charge density on the as-decorated GB is required, skipping the surface calculation required in the Rice-Wang framework. As such, it is for these reasons that we suggest that future research on segregant-induced effects on GB cohesion should instead focus on their atomic bonding and bond-breaking arguments overall.
\\\\
From our data, we can generate several rules-of-thumb regarding GB cohesion; first, solutes which are significantly larger than the host when placed in the bulk lattice generally always tend to be embrittling, e.g., the alkali metals, alkine earth metals, and noble gases. However, solutes of similar size to the host may not necessarily be beneficial, they can also be detrimental. The greatest strengtheners tend to be in the refractory metals, which are closest in size to the host metal, and the size of their strengthening effect increases with increasing row number. This coincides with the elements with the largest summed DDEC6 bond orders in the bulk [see Fig. \ref{SI:fig:BondOrderBulk_vs_Z} in the S.I.]. Solutes which are significantly smaller, e.g., the light elements, do not follow any specific rule, but those closer to the left of the periodic table tend to have the highest chance to be cohesion enhancing, whereas those to the right tend to be embrittling. Later period nonmetals are always embrittling. A visualisation of this data projected onto the periodic table is available in Figs. \ref{SI:fig:Ptable_segregation_engineering_ANSBO_bounded}, \ref{SI:fig:Ptable_segregation_engineering_Wsep_bounded} in the S.I.
\subsection{Outlook}
We now focus on the directions for future research. Recently, we, amongst others, have demonstrated that the effects of solutes and impurities in steels cannot be simply derived or extrapolated from their effects as single elements at GBs, but rather must be studied in their full chemical and spatial complexity \cite{ahmadianAluminumDepletionInduced2021, maiSegregationTransitionMetals2022, maiPhosphorusTransitionMetal2023, sakicInterplayAlloyingTramp2024, zhangGrainBoundaryAlloying2024}. Here, we further show that simplifications and commonly used approximations such as single-site/single cohesive effect/single model GBs cannot accurately represent segregation phenomena generally even in simple binary Fe-X systems, using ab-initio data across a \textit{set} of model GBs. Instead, the use of more complex models that incorporate both the spectral nature of segregation \cite{whiteSpectrumBindingEnergies1977} and the site-specific cohesive effects \cite{aksoySpectrumEmbrittlingPotencies2021} are necessary to accurately describe and understand segregation-induced cohesion phenomena overall. Extending prior investigations in the paramagnetic regime in both ferritic and austenitic GBs in steels  \cite{itoFirstprinciplesComputationalTensile2022, itoAnalysisGrainBoundary2022, hegdeAtomicRelaxationDefects2020} to incorporate these more complex frameworks are also crucial to furthering our understanding of segregation in steels.
\\\\
Wagih and Schuh have recently argued that studies utilising limited CSL models cannot be extrapolated to the full polycrystalline GB space \cite{wagihViewpointCanSymmetric2023}. Nevertheless, we have demonstrated that good agreement with experimentally measured segregation-related enrichment in a few specific cases (Nb, Mo enrichment in APT experiments \cite{maruyamaInteractionSoluteNiobium2003}), can be obtained regardless using such model GBs. Furthermore, we do not expect this to change the qualitative conclusions that we have drawn here with respect to the elemental trends of segregation binding or the spectral nature of \textit{cohesion}. However, we acknowledge that further, more comprehensive studies are required to investigate whether such agreement is merely serendipitous. While we have made efforts to cover the GB space to the extent computationally reasonable with DFT, future works, such as those involving machine-learning and machine-learned interatomic potentials, are in progress to address these realistic segregation scenarios in steels.
\\\\
The accompanying comprehensive data provided by our study should be of general interest to the community studying segregation to grain boundaries and defects overall. It should be particularly valuable to those developing magnetic machine learning interatomic potentials, serving as a chemically comprehensive ab-initio benchmark. We envision that this data should serve as a launching pad towards evermore complex and realistic studies of segregation phenomena.

\section{Conclusion}

Here, we present what is effectively a chemically-complete first-principles evaluation of elemental segregation and their effects on GB cohesion in ferritic Fe GBs. The generated data is used to derive a complete segregation engineering map for elemental segregation in ferromagnetic ferritic iron GBs. We highlight the data acquired for the lanthanides, which have not been studied to this extent before using ab-initio techniques.  The strongest segregation effects are driven exclusively by solute size effects. We propose that this site-volume/segregation binding strength relationship can be leveraged by experimental techniques to access site-level segregation favourability for specific elements. 
\\\\
The effects of segregated elements on GB cohesion were assessed using both bonding-based arguments, using DDEC6 area-normalised summed bond orders, as well as energetic arguments, in the form of results in Rice-Wang's theory of interfacial embrittlement. The most promising solutes for enhancing GB cohesion, from a purely elemental segregation perspective, were consistently identified to be mostly situated in the refractory metals block (e.g., Nb, Mo, Ta, W, Re, Os). Almost all other elements are expected to be neutral or embrittling in nature. The most detrimental solutes tend to be those with large size, i.e., the alkali metals, alkaline earths, noble gases, later period semimetals/nonmetals. Of engineering interest, in particular in the context of scrap recycling, are Sn, Sb and As, which all exhibit considerably strong segregation binding at GBs, and as such can be expected to outcompete beneficial alloying elements (Mo, W, Nb etc.). 
\\\\
Finally, and most importantly, using this wealth of data, we demonstrate that accounting for segregation spectra is not only critical in describing the amount of segregation, but also necessary for an accurate description of their induced effects on GB cohesion. One important consequence is that the effects of segregants on the cohesion of GBs cannot be assumed to be invariant over temperatures or GB types. This is illustrated using an example with segregating Mn. Since solutes exhibit differing segregation binding strengths to different sites, and different sites naturally enact distinct effects on GB cohesion, the effects of any element on GB cohesion in an alloy must also vary across different temperatures. Therefore, solutes which may appear embrittling or cohesion enhancing at specific conditions and GBs, may exhibit markedly different behaviours at others. Our results demonstrate the need for more complex models that incorporate both the spectral nature of segregation and the site-specific cohesive effects in order to accurately describe and understand segregation-induced cohesion phenomena. 

\section{Supplementary information and data availability}

The attached Supplementary Information PDF document contains data in the form of relaxed POSCAR files for the pure GBs (with vacuum slabs), the calculation information in the form of POTCARs used for each element, tabular data on the bulk-solute 127 Fe + 1 solute BCC cell study, the minimum segregation energies at each GB for each solute, associated data for the maximum segregation binding sites for each element in each GB, and plots of slices of the data referred to throughout the text (e.g. per-element/per-GB plots).
\\\\
The processed data, including all initial and final relaxed GB structures in pymatgen structure (v2023.11) \cite{ongPythonMaterialsGenomics2013} format, across all sites and elements, the calculation parameters (INCAR files), their magnetic moments, segregation energies, associated cohesive quantities (Wsep and ANSBO) are available in tabulated form as a pickled python pandas DataFrame \cite{thepandasdevelopmentteamPandasdevPandasPandas2024}, and CSV format (without the structures) at  \href{https://github.com/ligerzero-ai/FeGB_PtableSeg_FromFirstPrinciples_Data}{https://github.com/ligerzero-ai/FeGB\_PtableSeg\_FromFirstPrinciples\_Data}. The repository also contains the necessary tools to reproduce the environment to reproduce the pymatgen structures [i.e., \texttt{pip install .}] and the code used to generate the Figures. The corresponding raw calculation data (total energies, forces, stresses) will be uploaded to the same repository at a later date.
 
\section{Acknowledgements}
H.L. Mai, T. Hickel, J. Neugebauer acknowledge the support provided by the German Federal Ministry of Education and Research (BMBF) through the project Innovation-Platform MaterialDigital [Grant no. 13XP5094C]. T. Hickel, J. Neugebauer acknowledge financial support by the Deutsche Forschungsgemeinschaft (DFG) from CRC1394 “Structural and Chemical Atomic Complexity – From Defect Phase Diagrams to Material Properties”, project ID 409476157. This work was supported by computational resources provided by the Australian Government through the National Computational Infrastructure (Gadi) and the Pawsey Supercomputing Centre (Setonix) under the National Computational Merit Allocation Scheme. The Pawsey Supercomputing Centre is also supported by funding from the Government of Western Australia. Support and facilitation of our access to these compute resources from the Sydney Informatics Hub at the University of Sydney is gratefully acknowledged. S. Ringer acknowledges gratefully partial funding from the Australian Research Council. 

\bibliography{Manuscript-FePtable}
\end{document}